\documentclass[12pt]{article} %***
\usepackage[sectionbib]{natbib}
\usepackage{array,epsfig,fancyheadings,rotating}
\usepackage[]{hyperref}  %<----modified by Ivan
\usepackage{sectsty, secdot}
\usepackage{color}
\sectionfont{\fontsize{12}{14pt plus.8pt minus .6pt}\selectfont}
\renewcommand{\theequation}{\thesection\arabic{equation}}
\subsectionfont{\fontsize{12}{14pt plus.8pt minus .6pt}\selectfont}
\def\bSig\mathbf{\Sigma}

\newcommand{\bX}{\boldsymbol{X}}

\newcommand{\bZ}{\boldsymbol{Z}}

\newcommand{\bbeta}{\boldsymbol{\beta}}
\newcommand{\balpha}{\boldsymbol{\alpha}}

\newcommand{\trans}{^{\T}}
\makeatletter

\newcommand{\Rmnum}[1]{\expandafter\@slowromancap\romannumeral #1@}
\makeatother

\newcommand{\bP}{{\bf P}}
\newcommand{\bY}{{\bf Y}}
\newcommand{\bgamma}{\boldsymbol{\gamma}}
\newcommand{\bepsilon}{\boldsymbol{\epsilon}}
\newcommand{\T}{\!\top\!}

\usepackage{amsmath}
\usepackage{amssymb}
\usepackage{amsfonts}
\usepackage{multirow}
\usepackage{amsthm}

\allowdisplaybreaks

\theoremstyle{definition}

\begin{document}

%%%%%%%%%%%%%%%%%%%%%%%%%%%%%%%%%%%%%%%%%%%%%%%%%%%%%%%%%%%%%%%%%%%%%%%%%%%%%%%%%%%%%%%%%%%%%%%%%%%%%%%%%%%%%%%%%%%%%%%%%%%%
%%%%%%%%%%%%%%%%%%%%%%%%%%%%%%%%%%%%%%%%%%%%%%%%%%%%%%%%%%%%%%%%%%%%%%%%%%%%%%%%%%%%%%%%%%%%%%%%%%%%%%%%%%%%%%%%%%%%%%%%%%%%

\renewcommand{\baselinestretch}{1}

%\markright{ \hbox{\footnotesize\rm Statistica Sinica
%%{\footnotesize\bf 24} (201?), 000-000
%}\hfill\\[-13pt]
%\hbox{\footnotesize\rm
%%\href{http://dx.doi.org/10.5705/ss.20??.???}{doi:http://dx.doi.org/10.5705/ss.20??.???}
%}\hfill }
%
%\markboth{\hfill{\footnotesize\rm Huijuan Ma, Limin Peng and Haoda Fu} \hfill}
%{\hfill {\footnotesize\rm Quantile Regression of Latent Longitudinal Trajectory Features} \hfill}
%
%\renewcommand{\thefootnote}{}
%$\ $\par

%%%%%%%%%%%%%%%%%%%%%%%%%%%%%%%%%%%%%%%%%%%%%%%%%%%%%%%%%%%%%%%%%%%%%%%%%%%%%%%%%%%%%%%%%%%%%%%%%%%%%%%%%%%%%%%%%%%%%%%%%%%%

\fontsize{12}{14pt plus.8pt minus .6pt}\selectfont \vspace{0.8pc}
\centerline{\large\bf Quantile Regression of Latent Longitudinal}
\vspace{2pt} \centerline{\large\bf Trajectory Features}
\vspace{2pt} \centerline{Huijuan Ma$^1$, Limin Peng$^1$ and Haoda Fu$^2$} \vspace{2pt} \centerline{\it
       $^{1}$Department of Biostatistics and Bioinformatics, Emory University}  \vspace{2pt} \centerline{\it
       $^{2}$Eli Lilly and Company} \vspace{.35cm} \fontsize{9}{11.5pt plus.8pt minus
.6pt}\selectfont

%%%%%%%%%%%%%%%%%%%%%%%%%%%%%%%%%%%%%%%%%%%%%%%%%%%%%%%%%%%%%%%%%%%%%%%%%%%%%%%%%%%%%%%%%%%%%%%%%%%%%%%%%%%%%%%%%%%%%%%%%%%%

%\begin{quotation}
\noindent {\it Abstract:}
Quantile regression has demonstrated promising utility in longitudinal data analysis. Existing work is primarily focused on modeling cross-sectional outcomes, while  outcome trajectories often carry more substantive information in practice. In this work, we develop a trajectory quantile regression framework that is designed to robustly and flexibly investigate how latent individual trajectory features are related to observed subject characteristics. The proposed models are built under multilevel modeling with usual parametric assumptions lifted or relaxed. We derive our estimation procedure by novelly transforming the problem at hand to quantile regression with perturbed responses and adapting the bias correction technique for handling covariate measurement errors. We establish desirable asymptotic properties of the proposed estimator, including uniform consistency and weak convergence. Extensive simulation studies confirm the validity of the proposed method as well as its robustness.
An application to the DURABLE trial uncovers sensible scientific findings and illustrates the practical value of our proposals.

\vspace{9pt}
\par
%\end{quotation}\par

\def\thefigure{\arabic{figure}}
\def\thetable{\arabic{table}}

\renewcommand{\theequation}{\thesection.\arabic{equation}}

\fontsize{12}{14pt plus.8pt minus .6pt}\selectfont

\setcounter{equation}{0} %-1
\section{Introduction}

Longitudinal data, characterized by repeated measurements from the same subject, provide the essential platform for exploiting the temporal patterns of scientific outcomes. Such data frequently arise in biomedical research. A general account of methods for analyzing longitudinal data can be found in various texts \cite[among others]{Jones1993, Hand1996, Verbeke2000, Diggle2002, Fitzmaurice2004}.

Quantile regression \citep{Koenker1978}, given its robustness in handling skewed responses and flexibility in characterizing covariate effects, has demonstrated promising utility in longitudinal data analysis. A common way to formulate longitudinal quantile regression is to specify the outcome quantile at a given time point as a function of covariates, sharing a similar spirit with the generalized estimating equation (GEE) approach \citep{Liang1986}. A number of authors have studied such a marginal quantile regression model, including the GEE-type estimating equations and empirical-likelihood approaches \citep[among others]{Jung1996, He2003, Chen2004, Fu2012, Leng2014, Lu2015}. Extensions have also been proposed to address data complications such as dropouts and censoring \citep[for example]{Lipsitz1997, Wang2009,  Lee2013, Sun2016}. A more flexible modeling strategy for longitudinal quantile regression is to model outcome quantiles conditioning on covariates as well as fixed or random individual parameters that capture unobserved heterogeneity. Such conditional quantile regression models provide individual-specific interpretations and can be estimated through distribution-free or likelihood-based approaches \citep[among others]{Koenker2004, Harding2009, Galvao2010, Galvao2011}. %Despite the different quantile definitions,
It is worth noting that all these existing models are oriented to infer about the quantiles of the longitudinal outcome at given time points (i.e. cross-sectional quantiles).

\begin{figure}[h!]
%\captionsetup{width=0.8\textwidth}
\centering
\includegraphics[width=12cm, height=12cm]{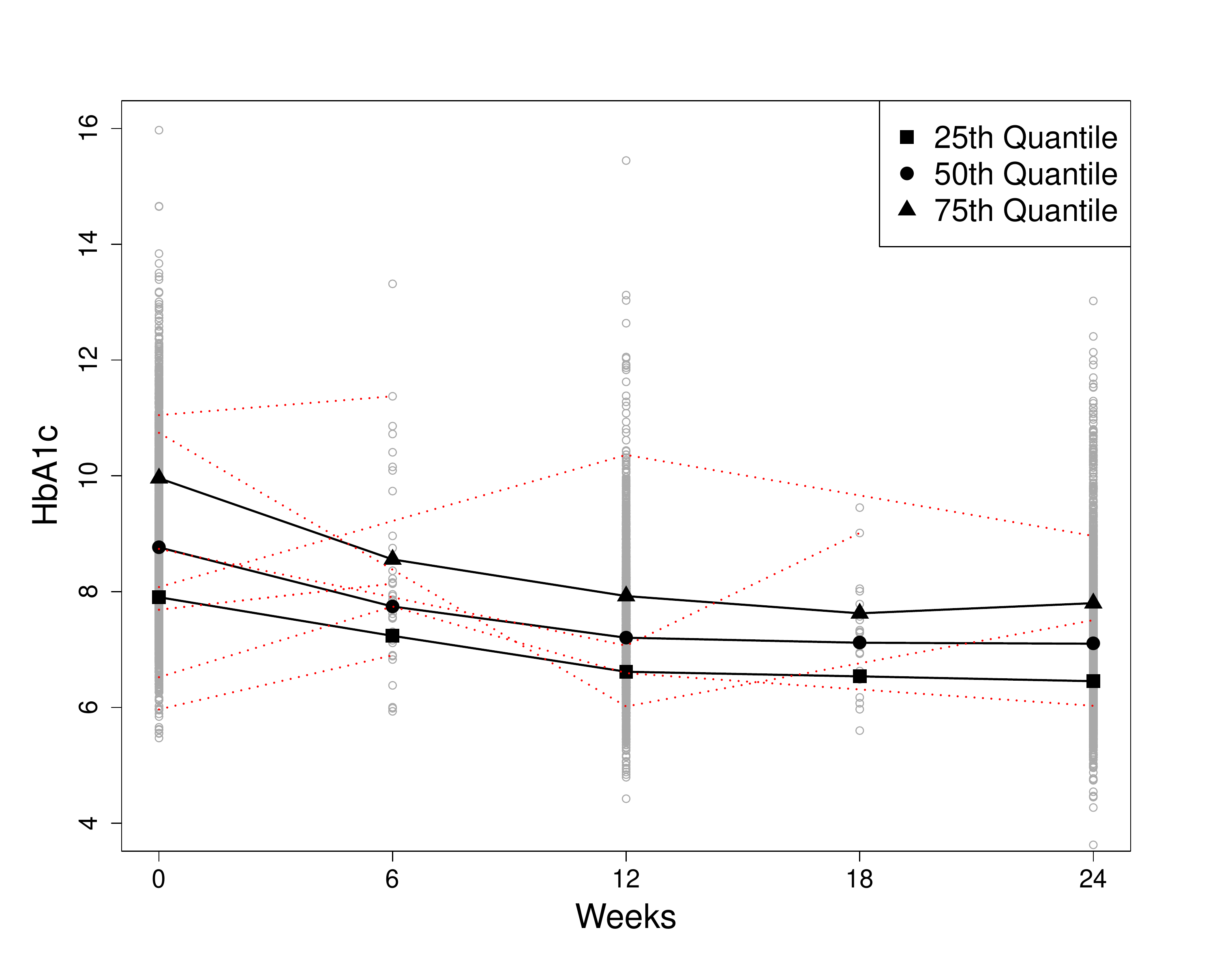}
\caption{DURABLE trial: HbA1c values over time. The solid black lines depict the 25th, 50th, and 75th quantiles of HbA1c at different time points; the red dotted lines examplify the HbA1c trajectories of 5 subjects. }
\label{fig0}
\end{figure}

In practice, there are many practical scenarios where the scientific interest pertains to the within-subject  trajectory of an outcome. % rather than its cross-sectional distribution/quantile functions.
To address such an interest, the current modeling of cross-sectional quantiles may not suffice because the changing pattern of cross-sectional quantiles over time often do not reflect the outcome changing pattern at the subject level. %This is because each subject may not necessarily stay at the same quantile level over time.
%As a simple example, the weight of a given infant may drop from the 90th percentile to the 50th percentile during the first year of life, while both the 50th and 90th population percentiles of baby weight still show an upward trend over time. %This work is motivated by many practical scenarios where the scientific interest pertains to the within-subject  trajectory of an outcome rather than its cross-sectional distribution/quantile functions.
For example, in the DURABLE trial \citep{Buse2009} that evaluated two starter insulin regimens in type 2 diabetes patients, how quickly HbA1c decreases over time within a patient may be a more substantive efficacy measure than the patient's HbA1c level at each follow-up visit.  This is because the trajectory of HbA1c would better inform us the patient's early response of treatment which is an effective indicator of the likely need for change in (or intensification of) therapy \citep{Fu2015}. As shown by Fig. \ref{fig0}, the 25th, 50th, and 75th cross-sectional quartiles of HbA1c are all decreasing with time (see black solid lines), while examining the within-subject data suggests that some subjects may have underlying HbA1c trajectories roughly unchanged or even increasing over time (see red dotted lines). These subjects correspond to weak responders to the assigned insulin treatment, which is of critical clinical importance but cannot be captured by evaluating the temporal trend of the cross-sectional quantiles. %Addressing an interest pertaining to the within-subject trajectory demands an alternative modeling perspective.

It is remarkable that the trajectory-related measures, such as the decreasing rate of within-subject HbA1c values, unlike the cross-sectional measurements, are not directly observable but may be accessible through trajectory modeling of HbA1c at the individual level.
To investigate whether and how a certain insulin treatment or individual characteristics lead to more rapid reductions of HbA1c, a natural approach is to consider multilevel modeling (MLM) \citep{Bryk2002}.  That is, one may assume a level-1 trajectory model for the repeated HbA1c measurements within each subject, and then use a level-2 model to link covariates
with the latent individual-specific features that capture the HbA1c reduction pattern based
on the level-1 model.  Traditional MLM for longitudinal data (or repeated measures) generally adopts i.i.d. normal random errors at every level of the model \citep[for example]{Snijders2002, Hedeker2006}. While this assumption helps ensure the model identifiability and facilitates likelihood-based inferences, it can
induce potentially stringent data constraints. Integrating quantile regression into repeated
measures MLM, for example in the level-2 model described above, can avoid some of unverifiable distributional assumptions. It can also offer the same flexibility to explore meaningful heterogeneous covariate effects as in the standard quantile regression. %, and result in a more robust and flexible model.

In this paper, we propose a new longitudinal quantile regression framework that serves to investigate individual trajectory features of longitudinal data under MLM. We shall refer the new framework to as longitudinal trajectory quantile regression. Our new model is easy to interpret, and complements the current marginal or conditional quantile regression models that are focused on the cross-sectional quantiles. As a proof of concept, we illustrate in Section 2 the proposed framework assuming a polynomial trajectory model and taking the outcome changing rate as the targeted trajectory feature.  The main thrust of our method is to deal with unobserved latent responses (e.g. decreasing rate of HbA1c) in the quantile regression context without fully parametric modeling. %without imposing parametric distributional assumptions.
Our key strategy is to map the latent quantile regression model to a quantile regression problem with observed but perturbed responses. As delineated in Section 3, we show that the bias caused by the response perturbance can be corrected through adapting the technique for handling covariate measurement errors \citep{Stefanski1990, Wang2012}.  Our estimation method permits a less restrictive trajectory model for the correlated within-subject measurements, for example, by allowing the error term to be non-normal. We carefully examine the developed inference procedures including the associated asymptotic properties and computational features; see Section 4. Our simulations reported in Section 5 confirm the validity of the proposed longitudinal trajectory quantile regression procedures and demonstrate their satisfactory performance  with finite sample sizes. The data application presented in Section 6 represents a novel secondary analysis of the DURABLE study, which reveals more robust and detailed heterogeneity patterns in longitudinal HbA1c outcomes.

\section{The Proposed Longitudinal Trajectory Quantile Regression Framework}

Consider a longitudinal study that consists of $n$ subjects. For subject $i$, let $Y_i(t)$ denote the outcome variable of interest.  We adopt a polynomial trajectory model for $Y_i(t)$ that takes the general form,
\begin{eqnarray} \label{model1}
Y_i(t) = f(t; \balpha_i) + \epsilon_i(t)\doteq \sum_{j=0}^k \alpha_{ij}t^{j}+\epsilon_i(t),
\end{eqnarray}
where $f(t; \balpha_i)$ is a $k$-th order polynomial function of time $t$, $\balpha_i=(\alpha_{i0}, \alpha_{i1}, \ldots, \alpha_{ik})\trans$ is an unknown {\em random} parameter vector, and $\epsilon_i(t)$ is a mean zero process independent of $\balpha_i$. Here the random parameters in $\balpha_i$ are not subject to the mean zero constraint and so they are general enough to capture both the fixed and random effects (of covariates) on $Y_i(t)$.
Under model \eqref{model1}, $f(t; \balpha_i)$ represents the underlying outcome trajectory of interest for subject $i$ determined by $\balpha_i$, and $\epsilon_i(t)$ may play a role like a random noise or measurement error.

{\it Remark 1:} It is important to point out that  the main interest under model \eqref{model1} is the underlying trajectory for subject $i$, $f(t;\balpha_i)$, and is {\em not} the conditional mean of $Y_i(t)$ given a covariate vector, say $\bX_i$, i.e. $E(Y_i(t)|\bX_i)\equiv E(f(t;\balpha_i)|\bX_i)$. This  distinguishes model \eqref{model1} from the traditional conditional mean modeling of longitudinal data.

The polynomial specification of $f(\cdot)$ can flexibly characterize various types of outcome temporal patterns. At the same time it provides the technical convenience that allows a meaningful trajectory-related feature of interest, denoted by $B_i$, be represented as a known parametric function of $\balpha_i$, $\phi(\balpha_i)$. With real data, the polynomial order $k$ may be determined by examining the  observed within-subject longitudinal measurements.  As motivated by the application to the DURABLE trial, we consider the special case where $\phi(\balpha_i)$ corresponds to the changing rate of the within-subject outcome trajectory  at a given time point $t_*$, which, under  model \eqref{model1}, takes the form
\begin{eqnarray}
B_i \doteq \phi(\balpha_i)= \frac{\partial f(t; \balpha_i) }{ \partial t}\bigg{|}_{t=t_*}=\sum_{j=1}^k j\alpha_{ij} t_*^{j-1}.
\end{eqnarray}
When $k=1$, $B_i$ is simply $\alpha_{i1}$, the random slope of the assumed linear trajectory. In practice, $t_*$  is often pre-specified per study protocol. Studying other forms of $\phi(\cdot)$, such as the area under curve of $Y_i(t)$ in a given time interval, can be carried out based on the same strategy presented in this work.

The focus of the proposed framework is to use quantile regression to permit a robust and comprehensive examination of the relationship between a latent outcome trajectory feature and the observed covariates. Let $\tilde{\bX}$ denote a $(p-1) \times 1$ vector of covariates,  $\bX=(1, \tilde{\bX}^{\T})^{\T}$, and let $B$ stand for the population analogue of $B_i$. The $\tau$th conditional quantile of $B$ given $\bX$ is defined as $Q_B (\tau | \bX)\doteq\inf\{b:\ \Pr(B\leq b|\bX)\geq \tau\}$. We assume the following quantile regression model:
\begin{eqnarray} \label{qrmodel}
Q_{B} ( \tau | \bX ) = \bX^{\T} \bbeta_0(\tau),
\end{eqnarray}
where $\tau\in(0,1)$ and $\bbeta_0(\tau) \in \mathcal{R}^p$ is an unknown coefficient vector. The coefficients in $\bbeta_0(\tau)$, as interpreted in the standard quantile regression \citep{Koenker2005}, reveal the change in the $\tau$-th quantile of $B$ per one unit covariate change. When all non-intercept coefficients in $\bbeta_0(\tau)$ are constant, model \eqref{qrmodel} reduces to a linear model where the regression coefficients represent the {\em location shift} to the distribution of $B$ resulted from one unit covariate change.

%{\it Remark 2:} A special case of the proposed models \eqref{model1} and \eqref{qrmodel} is the following linear mixed  model, which is commonly used for analyzing longitudinal data:
%$$
%Y_i(t)=\bX_i\trans \bb_1+a_{1, i}+(\bX_i\trans \bb_2+a_{2, i}) t+\epsilon_i(t),
%$$
%where $a_{1, i}$, $a_{2, i}$, and $\epsilon(t)$ are mean zero Gaussian and are independent. In this case, $B_i=\bX_i\trans \bb_2+a_{2, i}$ and $\bbeta_0(\tau)=(Q_{a_2}(\tau)+b_2^{(1)}, b_2^{(2)}.\ldots, b_2^{(p)})^T$, where $Q_{a_2}(\tau)$ denotes the $\tau$th quantile of $a_{2, i}$. Here and hereafter, the superscript $^{(j)}$ indicates the $j$th component of a vector.

Models \eqref{model1} and \eqref{qrmodel} together can be viewed as a multilevel model with level-1 units consisting of the repeated measurements for each subject and level-2 units corresponding to subjects. The multilevel model perspective enables us to employ quantile regression to investigate the determinants of a latent trajectory feature of interest.
%It also renders a simple interpretation of $\bbeta_0(\tau)$, which is the focus of the proposed longitudinal trajectory quantile regression framework.  Bearing in mind that quantile regression is a significant extension of traditional linear regression,
%{\color{red} As shown by Remark 2,}
The resulting multilevel model encompasses common repeated measures models such as linear mixed model with random intercept or slope. The quantile regression specification of the level-2 model avoids some of usual normality assumptions involved in the linear mixed models and hence leads to improved robustness.

\section{The Proposed Estimation}

%\lhead[\footnotesize\thepage\fancyplain{}\leftmark]{}\rhead[]{\fancyplain{}\rightmark\footnotesize\thepage}%Put this line in Page 2

In longitudinal studies, the outcome process $Y_i(t)$ is usually not continuously observed; rather $Y_i(t)$ is measured only at multiple discrete time points, say $t_{i1} < t_{i2} < \cdots < t_{i m_i}$, where $m_i$ is the total number of observations for subject $i$. Let $Y_{ij}$ denote the $Y(t)$ measured for subject $i$ at time point $t_{ij}$. The observed data consist of $\{ Y_{ij}, t_{ij}, \bX_i: j=1, \ldots, m_i;~ i=1, \ldots, n \}$. Our data scenario accommodates both regular and irregular time points for longitudinal measurements.

Model (\ref{model1}) implies that
%\begin{eqnarray}
$Y_{ij} = \alpha_{i0} + \alpha_{i1} t_{ij} + \alpha_{i2} t_{ij}^2 + \ldots + \alpha_{ik} t_{ij}^k + \epsilon_{ij}$
% \\      &\equiv&   \bZ_{ij} \balpha_i + \epsilon_{ij} ,        \label{remodel1}
%\end{eqnarray}
with $\epsilon_{ij} = \bepsilon_i(t_{ij})$ for $j=1, \ldots, m_i$ and $i=1, \ldots, n$.
This can further be expressed in a matrix form,
\begin{eqnarray} \label{remodel2}
\bY_i = \bZ_i \balpha_i + \boldsymbol{\epsilon}_i,
\end{eqnarray}
where $\bY_i=(Y_{i1},\ldots, Y_{i m_i})\trans$, $\bZ_i=(z_{p, q})$ is a $m_i\times (k+1)$ matrix with $z_{p, q}=t_{ip}^{q-1}$, and $\bepsilon_i=(\epsilon_{i1},\ldots, \epsilon_{im_i})\trans$, and $\balpha_i=(\alpha_{i0},\ldots, \alpha_{ik})\trans$.
%\begin{eqnarray*}
%\bY_i = \left( \begin{array}{c}
%              Y_{i1}  \\[1mm]
%              Y_{i2}  \\[1mm]
%              \vdots  \\[1mm]
%              Y_{i m_i}
%            \end{array}  \right),
%~~\bZ_i = \left( \begin{array}{cccc}
%              1 & t_{i1} & \cdots & t_{i1}^k \\[1mm]
%              1 & t_{i2} & \cdots & t_{i2}^k \\[1mm]
%              \vdots & \vdots & \ddots & \vdots \\[1mm]
%              1 & t_{im_i} & \cdots &t_{im_i}^k
%            \end{array}  \right),
%~~\bepsilon_i = \left( \begin{array}{c}
%              \epsilon_{i1}  \\[1mm]
%              \epsilon_{i2}  \\[1mm]
%              \vdots  \\[1mm]
%              \epsilon_{i m_i}
%            \end{array}  \right),
%~~\mbox{and}~~\balpha_i = \left( \begin{array}{c}
%              \alpha_{i0} \\[1mm]
%              \alpha_{i1} \\[1mm]
%              \vdots \\ [1mm]
%              \alpha_{ik}
%            \end{array}  \right).
%\end{eqnarray*}
Accordingly, we can write the latent response in model \eqref{qrmodel} as
%\begin{eqnarray}
$B_i %&\doteq& B_i(t_*) = \alpha_{i1} + 2 \alpha_{i2} t_* + \ldots + k \alpha_{ik} t_*^{k-1}   \nonumber \\
     %&=&
     =\bgamma^{\T} \balpha_i$,
%\end{eqnarray}
where $\bgamma=( 0, 1, 2 t_*, \ldots, k t_*^{k-1} )^{\T}$.

Estimating $\bbeta_0(\tau)$ would be a trivial quantile regression problem if $B_i$ ($i=1,\ldots, n$) were known. In that case, a well-studied estimator is given by
\begin{eqnarray}  \label{unest}
\mbox{argmin}_{\bbeta \in \mathcal{R}^p} \sum_{i=1}^n \rho_\tau(B_i - \bX_i^{\T}\bbeta),
\end{eqnarray}
where $\rho_\tau( v ) \doteq v \{ \tau - I( v < 0) \}$ is the quantile loss function (Koenker, 2005) and $I(\cdot)$ is the indicator function.
However, $B_i$ is a function of the latent parameter $\balpha_i$ in model (\ref{model1}), which is not observable. Therefore, the standard quantile regression method is not applicable to estimating $\bbeta_0(\tau)$.

Since $B_i$'s are not observable, a natural thought to handle this problem is to replace the $B_i$ in \eqref{unest} by its proxy  that is obtainable from the observed data. For example, an intuitive proxy of $B_i$ is given by $\hat B_i=\bgamma^{\T} \hat{\balpha}_i$, where $\hat\balpha_i\doteq (\bZ_i^{\T} \bZ_i)^{-1} \bZ_i^{\T} \bY_i$. %It is easy to see that $\hat B_i$ is unbiased about $B_i$, i.e. $E(\hat B_i)=E(B_i)$, despite the dependence among the longitudinal observations for subject $i$, $\{(Y_{ij}, t_{ij}),\ j=1,\ldots, m_i\}$. On the other hand, the discrepancy between $\hat B_i$ and $B_i$ (i.e. $\hat B_i-B_i$) is not (asymptotically) ignorable because the construction of $\hat B_i$ only uses $m_i$ observations, and $m_i$ is often bounded in practice. %$\hat B_i$ is not a consistent estimate for $B_i$. This is because the sample size involved in the least squares regression of $\bY_i$ on $\bZ_i$ is $m_i$, which is often bounded in practice and does not approach to $\infty$.
% is the least squares estimate obtained from fitting the trajectory model \eqref{model1} based on the longitudinal observations for subject $i$, $\{(Y_{ij}, t_{ij}),\ j=1,\ldots, m_i\}$, which are pretended to be independent.
However, simply  substituting $B_i$ with $\hat B_i$ in \eqref{unest}  to estimate $\bbeta_0(\tau)$ can lead to biased estimation. % questionable. %A critical concern on this approach is that  the difference between $B_i$ and $\hat B_i$ may not be negligible in the estimation of $\bbeta_0(\tau)$.
This is clearly shown  by the simulation studies presented in Section 4 and can be explained as follows. % confirm such a concern by demonstrating considerable bias resulted from this naive approach. %$\hat B_i$, the least square estimator of $B_i$ based on the longitudinal measurements of subject $i$.
%n the following we propose a procedure to properly address the issue resulted from replacing $B_i$ by $\hat B_i$ and develop  . %so that we can obtain a valid estimator of $\bbeta_0(\tau)$.

First, by simple algebra, we show that under model \eqref{model1}, %$\hat{\balpha}_i = (\bZ_i^{\T} \bZ_i)^{-1} \bZ_i^{\T} (\bZ_i \balpha_i + \bepsilon_i) = \balpha_i + (\bZ_i^{\T} \bZ_i)^{-1} \bZ_i^{\T} \bepsilon_i$. Further denote , then
\begin{eqnarray}   \label{Berr}
\hat{B}_i = B_i + \eta_i,
\end{eqnarray}
 where
$\eta_i = \bgamma^{\T} ( \bZ_i^{\T} \bZ_i )^{-1} \bZ_i^{\T} \bepsilon_i$.
We can see from \eqref{Berr} that $\hat B_i$ is an unbiased estimator of $B_i$, but its difference from $B_i$ may not be negligible because $m_i$, the number of longitudinal measurements within subject $i$ used to construct $\hat B_i$, is usually bounded. The error term $\eta_i$ in \eqref{Berr} has a similar flavor to covariate measurement errors concerned in literature \citep[See a summary]{Carroll2006}; both are mean zero but not negligible.  Treating the observable $\hat B_i$'s as the true responses in model \eqref{qrmodel} constitutes a quantile regression problem with perturbed responses. While several papers \citep[for example]{He2000, Wei2009, Wang2012, Wu2015} investigated quantile regression with covariate measurement errors, how to deal with response perturbance hasn't been studied. The response perturbance $\eta_i$ bears a notable distinction from a typical covariate measurement error; that is, the latter is usually assumed to be independent of covariates while $\eta_i$ is clearly covariate dependent and thus cannot be simply ignored in the regression setting.

In this work, we derive an appropriate bias-correction method to eliminate the bias caused by $\eta_i$.
Specifically, denote the data with observed responses as $\mathcal{O}_i=\{ \hat{B}_i, \bZ_i, \bX_i  \}$ and the data with unobserved responses as $\mathcal{U}_i=\{ B_i, \bZ_i, \bX_i  \}$.
%Suppose the variance of $\bepsilon_i$ depend on $\sigma^2, \bZ_i$ and $\bX_i$.
Following the corrected score argument in \cite{Stefanski1989} and \cite{Nakamura1990},  we may obtain a consistent estimator of $\bbeta_0$ by minimizing $\sum_{i=1}^n \rho_\tau^*(\mathcal{O}_i, \bbeta)$, where  $\rho_\tau^*(\cdot)$ satisfies
$
E[ \rho_\tau^*(\mathcal{O}_i, \bbeta) | B_i, \bZ_i, \bX_i ]=\rho_\tau(B_i-\bX_i^{\T}\bbeta).
$
However finding such a $\rho_\tau^*(\cdot)$ is difficult because $\rho_\tau( v )$ involves an indicator function and thus is not differentiable at $v=0$.
To overcome this difficulty, we propose to approximate $\rho_\tau(v)$ by a smooth function $\rho_{\tau, h}(v)$, where $h$ is a positive smoothing parameter. The strategy of smoothing $\rho_\tau(\cdot)$ was used  and justified in various quantile regression settings \citep[for example]{Wang2012, Wu2015}. Specifically, we shall adopt the smoothing scheme used in \cite{Horowitz1998}. Define $\rho_{\tau, h}(v) = v \{ \tau - 1 + K(v/h) \}$, where $K(\cdot)$ is a smooth function satisfying $\lim_{v \rightarrow -\infty} K(v) = 0$ and $\lim_{v \rightarrow \infty} K(v) =1$. It is clear that $1-K(v/h)$ converges to $I(v<0)$ as $h\rightarrow 0$ and hence $\rho_{\tau, h}(v) = v \{ \tau - 1 + K(v/h) \}$ approaches $\rho_\tau(v)$ as $h \rightarrow 0$.

Now our goal becomes finding a corrected quantile loss function $\rho_{\tau, h}^*(\mathcal{O}_i, \bbeta)$ such that
\begin{eqnarray}
E[ \rho_{\tau, h}^*(\mathcal{O}_i, \bbeta) | B_i, \bZ_i, \bX_i] &=& \rho_{\tau, h}(B_i-\bX_i^{\T}\bbeta).    \label{corrected}
                            % \\                               & \rightarrow & \rho_\tau( B_i-\bX_i^{\T}\bbeta)   \nonumber
\end{eqnarray}
Given $\rho_{\tau, h}(B_i-\bX_i^{\T}\bbeta)\rightarrow \rho_\tau( B_i-\bX_i^{\T}\bbeta)$ as $h \rightarrow 0$, an estimator of $\bbeta_0(\tau)$ may be given by
\begin{equation}
\tilde{\bbeta}_{n, h_n}(\tau)= \mbox{argmin}_{\bbeta \in \mathcal{R}^p} \sum_{i=1}^n \rho_{\tau, h_n}^*(\mathcal{O}_i, \bbeta), \label{corrected2}
\end{equation}
with $h_n\rightarrow 0$ as $n\rightarrow\infty$.

Since the distribution of $\mathcal{O}_i$ given $(B_i, \bZ_i, \bX_i)$ is determined by the distribution of $\bepsilon_i$,  \eqref{corrected} suggests that the form of $\rho_{\tau, h}^*(\cdot)$ depends on the distribution of $\bepsilon_i$.
In the following two subsections, we shall construct $\rho_{\tau, h}^*(\cdot)$ for the cases where $\epsilon_{ij}$ follows the normal and the Laplace distributions respectively. The proposed methods then permit either normally distributed or heavy-tailed errors in the adopted trajectory model \eqref{model1}. We also evaluate the robustness of our method to misspecification of the distribution of $\bepsilon_i$ via simulations.

%\vspace*{-24pt}
\subsection{Normal trajectory random error}

%\lhead[\footnotesize\thepage\fancyplain{}\leftmark]{}\rhead[]{\fancyplain{}\rightmark\footnotesize\thepage}%Put this line in Page 2

%The assumption that the error follows a normal distribution is extensively adopted in statistical literature, including the linear regression model and the measurement error model (Fuller, 1987; Carroll et al, 2006).
Assume that $\{ \epsilon_{ij}, j=1, \ldots, m_i \}$ are independent, and  $\epsilon_{ij}|(\bX_i, \bZ_i) \sim  N(0, \delta(\bX_i, \bZ_i) \sigma^2  )$, where $\delta(\cdot)$ is a known positive scalar function and $\sigma^2>0$. In this case, the within-subject correlations in $Y_{ij}$'s are captured by the subject-specific latent random parameter $\balpha_i$. Then %$\boldsymbol{\epsilon}_i \sim N(0, \delta(\bX_i, \bZ_i) \sigma^2 I_{m_i} )$ is a $m_i$--dimensional normal random vector. Furthermore,
$
(\bZ_i^{\T} \bZ_i)^{-1} \bZ_i^{\T} \bepsilon_i \sim N(0, \delta(\bX_i, \bZ_i) \sigma^2 \{ \bZ_i^{\T} \bZ_i\}^{-1} )
$
and $\eta_i=\hat{B}_i-B_i \sim N(0, \sigma^2 D_i)$, where $D_i = \delta(\bX_i, \bZ_i) \bgamma^{\T} \{ \bZ_i^{\T} \bZ_i\}^{-1} \bgamma$. This implies
\begin{equation}
D_i^{-1/2} ( \hat{B}_i - \bX_i^{\T} \bbeta ) | (B_i, \bZ_i, \bX_i) \sim N \left( D_i^{-1/2} (B_i-\bX_i^{\T} \bbeta), \sigma^2 \right).
\label{normal1}
\end{equation}

Based on \eqref{normal1}, we can derive a corrected quantile loss function $\rho_{\tau, h}^*(\cdot)$  by employing the result established in \cite{Stefanski1995}. That is,  given a sufficiently smooth function $g(\cdot)$, and independent random variables, $U \sim N(\mu, \sigma^2)$ and $V \sim N(0, 1)$, it holds that $ E[ E\{ g( U + i \sigma V)| U \}] = g (\mu)$, where $i=\sqrt{-1}$. By Taylor expansion and the moment expression of standard normal distribution,
$
E\{ g( U + i \sigma V )| U \} = \sum_{m=0}^\infty g^{(2m)}(U) \frac{(-\sigma^2)^m}{2^m m!}.
$
Therefore
\begin{equation}
E\left\{\sum_{m=0}^\infty g^{(2m)}(U) \frac{(-\sigma^2)^m}{2^m m!}\right\}=g(\mu).
\label{normal2}
\end{equation}

The fact \eqref{normal2} provides the key insight on how to find the corrected quantile loss function.
Define $\hat{\xi}_i= D_i^{-1/2}(\hat{B}_i-\bX_i^{\T} \bbeta)$ and $\xi_i = D_i^{-1/2}(B_i-\bX_i^{\T}\bbeta)$. Viewing $\hat{\xi}_i$ as the $U$ in \eqref{normal2}, we obtain from \eqref{normal1} and \eqref{normal2} that
$
E[ \rho_{\tau, h}^{N*}(\mathcal{O}_i, \bbeta, \sigma^2) |B_i, \bX_i, \bZ_i]= \rho_{\tau, h} \left( \xi_i \right) %\rightarrow \rho_{\tau}( \xi_i )
$,
where
$
\rho_{\tau, h}^{N*}(\mathcal{O}_i, \bbeta, \sigma^2) %= \rho_{\tau, h}^{N*}( \hat{\xi}_i, \sigma^2)
 = \sum_{m=0}^\infty \rho_{\tau, h}^{(2m)} \left( \hat{\xi}_i \right)  \frac{(-\sigma^2)^m}{2^m m!}.
$
Note $\rho_{\tau}(\xi_i)=D_i^{-1/2} \rho_{\tau}(B_i-\bX_i\trans\bbeta)$ and $D_i>0$ given $\bZ_i\trans\bZ_i>0$. Following the argument for \eqref{corrected2},
$\rho_{\tau, h}^{N*}(\mathcal{O}_i, \bbeta, \sigma^2)$ may serve as a corrected quantile loss function if $\sigma^2$ is known. When $\sigma^2$ is unknown, we propose to employ  $\rho_{\tau, h}^{N*}(\mathcal{O}_i, \bbeta, \hat\sigma^2)$, where  $\hat\sigma^2$ is a reasonable estimator of $\sigma^2$ discussed in Section 3.3.

To compute $\rho_{\tau, h}^{N*}(\cdot)$, it is easy to calculate that $\rho_{\tau, h}^{(0)}(v)=\rho_{\tau, h}(v)$, $\rho_{\tau, h}^{(1)}(v) = \tau - 1 + \{ v K(v/h) \}^{(1)}$ and $\rho_{\tau, h}^{(j)}(v) = \{ v K(v/h) \}^{(j)}$ for $j \geq 2$, where
$
\left\{ v K \left( \frac{v}{h} \right)  \right\}^{(j)}  =  \frac{j}{h^{j-1}}  K^{(j-1)}  \left( \frac{v}{h} \right)+ \frac{v}{h^j}  K^{(j)}  \left( \frac{v}{h} \right), ~~j=1,2,\ldots .
$
Here we choose $K(\cdot)$  as an infinitely smooth function, such as the distribution function of $N(0, 1)$, which is adopted in our simulation studies. Solving the minimization of $\sum_{i=1}^n   \sum_{m=0}^\infty\rho_{\tau, h}^{(2m)} \left( \hat{\xi}_i \right)  \frac{(-\sigma^2)^m}{2^m m!}$ however involves an infinite series, and thus may not be feasible.  Following the practical recommendation of \cite{Stefanski1989} and \cite{Wu2015}, we shall keep the first two summands in $\rho_{\tau, h}^{N*}(\cdot) $ as an approximation to $\rho_{\tau, h}^{N*}(\cdot)$, which is found to be adequate in our simulation studies.

%\lhead[\footnotesize\thepage\fancyplain{}\leftmark]{}\rhead[]{\fancyplain{}\rightmark\footnotesize\thepage}%Put this line in Page 2

\subsection{Laplace Trajectory Random Error}

%\lhead[\footnotesize\thepage\fancyplain{}\leftmark]{}\rhead[]{\fancyplain{}\rightmark\footnotesize\thepage}%Put this line in Page 2

%The Laplace distribution is widely used in practice for modelling data with tails heavier than normal.
%Laplace errors are commonly discussed in measurement error literature, such as Stefanski and Carroll (1990), Hong and Tamer (2003), Wang et al.~(2012) and Wu et al.~(2016).
In this subsection, we consider the situation where the errors $\{\epsilon_{ij}, j=1,\ldots, m_i; i=1, \ldots, n\}$ follow a {univariate Laplace} distribution.
We adopt the classical definitions of univariate and multivariate Laplace distributions \citep[see Chapter 6]{Kotz2001}, some related properties of which can be found in Lemma 1 of \cite{Wang2012}.

Assume that $\{\epsilon_{ij},\ j=1,\ldots, m_i\}$ are independent and $\epsilon_{ij} |(\bX_i, \bZ_i) \sim { L}(0, \delta(\bX_i, \bZ_i) \sigma^2)$. Here $\delta(\cdot)$ is a known positive scalar function and $\sigma^2>0$. It is easy to see that
\begin{equation}
D_i^{-1/2} ( \hat{B}_i - \bX_i^{\T} \bbeta ) | (B_i, \bZ_i, \bX_i) \sim {L} \left( D_i^{-1/2} ( B_i-\bX_i^{\T} \bbeta), \sigma^2 \right).\label{laplace1}
\end{equation}
By Lemma 2 of \cite{Wang2012}, for a random variable $U$ following the univariate Laplace distribution ${L }(\mu, \sigma^2)$ and  a twice-differentiable function $g(\cdot)$,
\begin{equation}
E \left\{ g(U) -   \frac{\sigma^2}{2} g^{(2)}(U) \right\} = g( \mu ).
\label{laplace2}
\end{equation}
% where $g^{(2)}(\cdot)$ is the second derivative of $g$.
Choose $K(\cdot)$ be a twice-differentiable function, % which leads to $\rho_{\tau, h}(v) = v \{ \tau - 1 + K(v/h) \}$ that is twice-differentiable,
and denote
%\begin{eqnarray*}
$\rho_{\tau, h}^{L*}(v, \sigma^2) = \rho_{\tau, h}(v)-\frac{\sigma^2}{2}  \frac{ \partial^2 \rho_{\tau, h}(v)}{\partial v^2}
                  = v  \left\{ \tau -1 + K \left( \frac{v}{h} \right) \right\} - \frac{\sigma^2}{2}
                      \left\{ \frac{2}{h} K^{(1)} \left( \frac{v}{h} \right) + \frac{v}{h^2} K^{(2)} \left( \frac{v}{h} \right)   \right\}.
                  $
%\end{eqnarray*}
We then obtain from \eqref{laplace1} and \eqref{laplace2} that
$
E[ \rho_{\tau, h}^{L*}( \hat{\xi}_i, \sigma^2) |B_i, \bX_i, \bZ_i] = \rho_{\tau, h}(\xi_i),
$
where $\hat{\xi}_i = D_i^{-1/2} ( \hat{B}_i - \bX_i^{\T} \bbeta )$ and $\xi_i = D_i^{-1/2} ( B_i - \bX_i^{\T} \bbeta )$ are the same as those defined in Section 3.1.
A corrected quantile loss function for the Laplace error case (with known $\sigma^2$)  is thus given by
\begin{equation}
\rho_{\tau, h}^{L*}(\mathcal{O}_i, \bbeta, \sigma^2) = \rho_{\tau, h}^{L*}(\hat{\xi}_i, \sigma^2)
\hat{\xi}_i \left\{ \tau-1 + K \left( \frac{\hat{\xi}_i}{h} \right) \right\} - \frac{ \sigma^2 }{2}
                           \left\{ \frac{2}{h} K^{(1)} \left( \frac{\hat{\xi}_i}{h} \right) + \frac{\hat{\xi}_i}{h^2} K^{(2)} \left( \frac{\hat{\xi}_i}{h} \right)   \right\}.
\end{equation}

It is interesting to note that the corrected  quantile loss function for the Laplace error case is exactly the same as the approximate corrected loss function derived for the normal error case (that uses the first two summands in the infinite series). This fact enables a unified corrected quantile loss function for estimating $\bbeta_0(\tau)$ in the presence of normal or Laplace trajectory errors. When $\sigma^2$ are unknown, we shall plug in the estimator of $\sigma^2$ presented in Section 3.3.

%{\it Remark 3:} As shown in Sections 3.1 and 3.2, our method assumes the known {\em type} of the trajectory random error distribution but allows for a flexible specification.  For example, the incorporation of $\delta(\bX_i, \bZ_i)$ into the assumed normal or Laplace distributions permits the presence of data heteroscedasticity.  Our simulation studies in Section 5 suggest very promising robust performance of the proposed method to the misspecification of trajectory random error distribution type.

\subsection{Estimation of $\sigma^2$}

%Thus far we have described the proposed method under the assumption that the variance of $\epsilon_{ij}$ depend on $\sigma^2$ and the subject specific characteristics through $\delta(\bX_i, \bZ_i)$, and they are known.
%But in practice, $\sigma^2$ is not always known.
In this subsection, we discuss how to estimate $\sigma^2$.
Define
%\begin{eqnarray*}
$\tilde{\bY}_i ={\bY_i}/{\sqrt{\delta(\bX_i, \bZ_i)} }$, $\tilde{\bZ}_i ={\bZ_i}/{\sqrt{\delta(\bX_i, \bZ_i)} }$, and $\tilde{\bepsilon}_i ={\bepsilon_i}/{\sqrt{\delta(\bX_i, \bZ_i)} }$.
%\end{eqnarray*}
Write $\tilde{\bepsilon}_i = ( \tilde{\epsilon}_{i1}, \tilde{\epsilon}_{i2}, \ldots, \tilde{\epsilon}_{i m_i} )^{\T}$. By the definition, $\{ \tilde{\epsilon}_{ij}, j=1, \ldots, m_i; ~i=1, \ldots, n \}$ are independent and identically distributed (i.i.d.) with $E(\tilde{\epsilon}_{ij})=0$ and $\mbox{var}(\tilde{\epsilon}_{ij})=\sigma^2$.

Suppose their exists a constant $M$ such that $m_i < M < \infty$.
Following the idea of \cite{Sun2007}, we estimate $\sigma^2$ based on the residuals,
$\hat{ \bepsilon }_i = \tilde{\bY}_i - \tilde{\bZ}_i \hat{\balpha}_i = \tilde{\bY}_i - \tilde{\bZ}_i ( \tilde{\bZ}_i^{\T} \tilde{\bZ}_i)^{-1} \tilde{\bZ}_i^{\T} \tilde{\bY}_i = (I_{m_i}-\bP_i)\tilde{\bY}_i$, $i=1,\ldots, n$, where $\bP_i = \tilde{\bZ}_i ( \tilde{\bZ}_i^{\T} \tilde{\bZ}_i)^{-1} \tilde{\bZ}_i^{\T}$.
Let
$
\mbox{RSS}_i = \hat{\bepsilon}_i^{\T} \hat{\bepsilon}_i = \tilde{\bY}_i^{\T} (I_{m_i}-\bP_i) \tilde{\bY}_i$, and $q=k+1$, which correspond to the number of columns of $\tilde{\bZ}_i$.
%the synthetic degrees of freedom of $\mbox{RSS}_i$ is $m_i-q$.
Pooling all $\mbox{RSS}_i\ (i=1, \ldots, n)$ together naturally leads to an estimator of $\sigma^2$,
$
\hat{\sigma}^2 = \frac{1}{N-q n} \sum_{i=1}^n \mbox{RSS}_i,
$
where $N=\sum_{i=1}^n m_i$. By Lemma 1 of the Appendix, $\hat{\sigma}^2$ is consistent and asymptotically normal.

With the consistent estimator, $\hat{\sigma}^2$,  our proposed estimator $\hat{\bbeta}(\tau)$  when $\sigma^2$ is unknown is given by
$
\hat{\bbeta}_{n, h_n}(\tau) = \mbox{argmin}_{\bbeta \in \mathcal{B}} \sum_{i=1}^n   \rho_{\tau, h_n}^* \left( \mathcal{O}_i, \bbeta, \hat{\sigma}^2 \right).
$
Here and in the sequel, the notation $\rho_{\tau, h}^*\left( \mathcal{O}_i, \bbeta\right)$ in \eqref{corrected} is expanded to incorporate the additional argument from $\sigma^2$ or $\hat\sigma^2$.  The $\rho_{\tau, h_n}^*(\cdot)$ above stands for either $\rho_{\tau, h}^{N*}(\cdot)$ or $\rho_{\tau, h}^{L*}(\cdot)$.

\subsection{Choose the smoothing parameter $h$}

Motivated by the work of \cite{Delaigle2008} and \cite{Wang2012}, we propose a modified simulation-extrapolation-type strategy to choose the smoothing parameter $h$.

For notation simplicity and clarity, in this subsection we drop the $\tau$ in $\bbeta(\tau)$ and use $\hat{\bbeta}(h)$ to denote the estimator associated with smoothing parameter $h$. Define $M(h)=E[ \{ \hat{\bbeta}(h)-\bbeta_0 \}^{\T} \Omega^{-1} \{ \hat{\bbeta}(h)-\bbeta_0 \}]$, where $\Omega=\mbox{cov}\{ \hat{\bbeta}(h) \}$. It is natural to define the optimal smoothing parameter as $h_0= \mbox{argmin}_h M(h)$. However, since $M(h)$ depends on unknown $\bbeta_0$, the minimization of $M(h)$ cannot be carried out directly in practice. Instead, we propose to approximate $h_0$ through simulations.

Specifically, let $\{ \eta_{c1}^*, \ldots, \eta_{cn}^*  \}$ and $\{ \eta_{c1}^{**}, \ldots, \eta_{cn}^{**} \}$ denote two sequences of i.i.d. random variables from $N(0, \sigma^2 D_i)$ if the normal trajectory error is assumed or from $L(0, \sigma^2 D_i)$ if the Laplace trajectory error is assumed, where $c=1, \ldots, n_c$. When $\sigma^2$ is unknown, we replace it  by $\hat\sigma^2$.
Let
$
\hat{B}_{ci}^*=\hat{B}_i+\eta_{ci}^*$ and $\hat{B}_{ci}^{**}=\hat{B}_{ci}^*+\eta_{ci}^{**}$.
Let $\bbeta_c^*(h)$ and $\bbeta_c^{**}(h)$ be the proposed estimators obtained from samples $\{(\hat{B}_{ci}^*, \bX_i)\}_{i=1}^n$ and $\{(\hat{B}_{ci}^{**}, \bX_i)\}_{i=1}^n$, respectively. Define
$
M_1(h)= n_c^{-1} \sum_{c=1}^{n_c} \{ \bbeta_c^*(h)-\hat{\bbeta}(h)  \}^{\T} (S^*)^{-1} \{ \bbeta_c^*(h)-\hat{\bbeta}(h) \}$ and
$M_2(h)= n_c^{-1} \sum_{c=1}^{n_c} \{ \bbeta_c^{**}(h)-\bbeta^*(h)  \}^{\T} (S^{**})^{-1} \{ \bbeta_c^{**}(h)-\bbeta^*(h)  \}$,
where $S^*$ and $S^{**}$ are the sample covariance matrices of $\{\bbeta_c^*(h)-\hat{\bbeta}(h): c=1, \ldots, n_c \}$
and $\{ \bbeta_c^{**}(h)-\bbeta^*(h): c=1, \ldots, n_c \}$, respectively.
Let $\hat{h}_1 = \mbox{argmin}_h M_1(h)$ and $\hat{h}_2 = \mbox{argmin}_h M_2(h)$. Since $\hat{B}_{ci}^{**}$ measures $\hat{B}_{ci}^*$ in the same way that $\hat{B}_{ci}^*$ measures $B_i$, it is reasonable to expect that the relationship between $\hat{h}_2$ and $\hat{h}_1$ is similar to that between $\hat{h}_1$ and $h_0$. Therefore, back extrapolation can be used to approximate $h_0$. In our implementation, we use the linear extrapolation from the pair $(\log \hat{h}_1, \log \hat{h}_2)$ and define the second-order approximation to $h_0$ as $\hat{h}_0=\hat{h}_1^2/\hat{h}_2$.

\subsection{Computational considerations}

The proposed estimator of $\bbeta_0(\tau)$ is obtained through minimizing $\sum_{i=1}^n   \rho_{\tau, h_n}^* \left( \mathcal{O}_i, \bbeta, {\sigma}^2 \right)$ (if $\sigma^2$ is known) or $\sum_{i=1}^n   \rho_{\tau, h_n}^* \left( \mathcal{O}_i, \bbeta, \hat{\sigma}^2 \right)$ (if $\sigma^2$ is unknown). In either case, the objective function (multiplied by $1/n$) is smooth and  converges to $E\{D_i^{-1/2}\rho(B_i-\bX_i\trans\bbeta)\}$, a convex function of $\bbeta$, as $n\rightarrow\infty$. This means, when $n$ is large, the corrected objective function, $\sum_{i=1}^n   \rho_{\tau, h_n}^* \left( \mathcal{O}_i, \bbeta, {\sigma}^2 \right)$, should be close to a smooth convex function, and thus usual optimization algorithms can work adequately for finding $\hat\bbeta(\tau)$.
When $n$ is not large enough, setting $h$ too small may yield a jagged corrected objective function that presents many multiple local minimizers. Similar observations were made in others' work, for example, \cite{Stefanski1985, Stefanski1987}, \cite{Nakamura1990}, and \cite{Wang2012}.  In that case, one may consider increasing $h$; doing so often leads to a less rough objective function better conforming to a convex pattern. In our implementation, we locate the minimizer of the corrected objective function by using the R function $optim()$. We set the initial value for searching the minimizer as the naive estimator obtained from qualtile regressing $\hat{B}_i$ on $\bX_i$.

%\lhead[\footnotesize\thepage\fancyplain{}\leftmark]{}\rhead[]{\fancyplain{}\rightmark\footnotesize\thepage}%Put this line in Page 2

%\section{Asymptotics and Inferences}

%\vspace{-10pt}
\section{ {Inferences } }  \label{varest}

Inferences about $\hat{\bbeta}(\tau)$ are important in applications. {In the Supplementary Materials, we present the large sample properties of the proposed estimator, including uniform consistency and weak convergence to a Gaussian process; see Section S1 of the Supplementary Materials.} By our theory, the limiting variance of $n^{1/2} \{ \hat{\bbeta}(\tau)- \bbeta_0(\tau)\}$ involves complex quantities, estimation of which may be unstable with small or moderate sample sizes. In this subsection, we propose to use a simple resampling approach to estimating the asymptotic variance. Our procedure adapts the perturbing technique proposed by \cite{Jin2001} to the estimation setting considered here. % and then used by Li and Wang (2008) for longitudinal data to the situation consider here.

Let $\omega_1$, $\omega_2$, $\ldots$, $\omega_n$ be independent variates from a nonnegative known distribution with mean 1 and variance 1, for example, $Exponential(1)$.
With the data fixed at the observed values, we repeatedly generate the variates $\{ \omega_1, \omega_2, \ldots, \omega_n \}$ and obtain a large number of realizations of
$
\sigma_*^2 = \frac{(N-qn)^{-1} \sum_{i=1}^n \omega_i~ \mbox{RSS}_i }{ n^{-1} \sum_{i=1}^n \omega_i},
$
and
%\begin{eqnarray}  \label{bestar}
$\bbeta^*(\tau) = \mbox{argmin}_{\bbeta \in \mathcal{B}} \sum_{i=1}^n  \omega_i ~  \rho_{\tau, h}^* \left( \mathcal{O}_i, \bbeta, \sigma_*^2 \right)$,
%\end{eqnarray}
denoted by $\{  \bbeta_r^*(\tau) \}_{r=1}^{n_b}$. In the Supplementary Materials, we show that the conditional distribution of $n^{1/2} \{ \bbeta^*(\tau) - \hat{\bbeta}(\tau) \}$ given the observed data is asymptotically the same as the unconditional distribution of $n^{1/2} \{ \hat{\bbeta}(\tau) - \bbeta_0(\tau) \}$. Hence, the variance of $\hat{\bbeta}(\tau)$ can be estimated by the sample variance of $\{ \bbeta_r^*(\tau) \}_{r=1}^{n_b}$. The confidence intervals for $\bbeta(\tau)$ can be constructed using the normal approximation or by referring to the empirical percentiles of $\bbeta^*(\tau)$.

When $\sigma^2$ is known, we can skip the resampling step for $\sigma_*^2$ and simply replace the $\sigma_*^2$ in {obtaining $\bbeta^*(\tau)$} by $\sigma^2$. The justification of the presented resampling-based inference procedure is provided in the Supplementary Materials (see Section S2)

In  addition, we may conduct second-stage inference procedures to render useful summaries of $\bbeta_0(\tau)$ and explore the varying patterns of $\bbeta_0(\tau)$ over $\tau$, following the lines of \cite{Peng2009}; please see the  details in Section S3 of the Supplementary Materials.

\section{Simulation Studies}

We conduct extensive simulations to investigate the performance of the proposed method with finite samples.
We generate the longitudinal data based on models (\ref{remodel2}) and (\ref{qrmodel}). Specifically, we generate the longitudinal observations $Y_{ij}$ as
%\begin{eqnarray} \label{simumodel1}
$Y_{ij} = a_i + b_i t_{ij}+ \epsilon_{ij}$, $j=1, \ldots, m_i$; $i=1, \ldots, n$,
%\end{eqnarray}
where $a_i$ is the random intercept, following the exponential distribution, $Exp(0.8)$, $b_i$ is the random slope generated from the following linear model with heteroscedastic errors,
%\begin{eqnarray}
$b_i = 2 + X_{1i} + X_{2i} + (0.1+X_{1i} + X_{2i})e_i$. %. \label{simumodel1_2}
%\end{eqnarray}
%For model \eqref{simumodel1},
We generate observation time points,
$\{t_{ij}, j=1, \ldots, m_i \}$, from a Poisson process with intensity 0.8 (i.e.  $t_{ij}$ is the sum of $j$ independent exponential variables $Exp(0.8)$). We specify
$\{m_i,\ \ i=1,\ldots, n\}$ as i.i.d. random variables that are defined as  the integer part of $4+U_i$, where $U_i$ follows the uniform distribution, $Unif(0, 6)$.
%For model \eqref{simumodel1_2},
We generate $X_{1i}$ from the uniform distribution, $Unif(0, 0.5)$, $X_{2i}$  from the Binomial distribution, $Ber(0.5)$,  and $e_i$ from the standard Normal distribution, $N(0,1)$. With the random slope $b_i$ being the trajectory feature of interest $B_i$, it is easy to see that
%show that {model \eqref{simumodel1_2} implies (delete because ??)}
$Q_{B_i}(\tau|\bX_i)=\beta_{0}(\tau)+\beta_{1}(\tau) X_{1i} + \beta_{2}(\tau) X_{2i}$ with $\beta_{0}(\tau) = 2 + 0.1 Q_e(\tau)$ and $\beta_{1}(\tau) = \beta_{2}(\tau)=1+Q_e(\tau)$, where $Q_e(\tau)$ is the $\tau$th quantile of $e_i$.

We consider four different configurations for the error term in the trajectory model, $\epsilon_{ij}$.  We examine the cases with heavy-tailed trajectory errors and the cases where the trajectory errors depend on the covariates $\bX_i$ . The specific configurations are listed as follows:

Case 1: ~~$\epsilon_{ij} \sim { L(0, 1)}$,  the univariate asymmetric Laplace distribution.

Case 2: ~~$\epsilon_{ij} \sim N(0, 1)$, the standard Normal distribution.

Case 3: ~~$\epsilon_{ij} = \epsilon_{ij}^*/(1+X_{i1})$ with $\epsilon_{ij}^* \sim { L(0, 1)}$.

Case 4: ~~$\epsilon_{ij} = \epsilon_{ij}^*/(1+X_{i1})$ with $\epsilon_{ij}^* \sim N(0, 1)$.\\
In Case 3 and Case 4, $(1+X_{i1})$ serves as the $\delta(\bX_i, \bZ_i)$ in the trajectory error distributions discussed in Sections 3.1 and 3.2.

Under each configuration, we simulate 1000 datasets of sample size $200$ or $500$. We apply the proposed method to each simulated dataset, examining the covariate effects on the 10th, 20th, $\ldots$, 90th quantiles of $B_i$. We compare the proposed estimator with the naive estimator, which is obtained by directly quantile regressing $\{\hat{B}_i\}_{i=1}^n $ on $\{\bX_i=(X_{1i}, X_{2i})^{\T}\}_{i=1}^n$.
Our proposed estimator %, which minimizes the objective function coupled with treating the error as Laplace,
is obtained through the $optim()$ function in R with the initial value set as he naive estimator. The bandwidth $h$ is chosen as 0.8 and the smoothing function $K(\cdot)$ is chosen as the standard normal distribution function. We use the resampling method presented in Section \ref{varest} to estimate the asymptotic variance of the proposed estimator, choosing $n_b=200$ and generating $\{ \omega_1, \ldots, \omega_n \}$ from the exponential distribution $Exp(1)$.

\begin{figure}[h!]
\centering
\includegraphics[width=16cm, height=16cm]{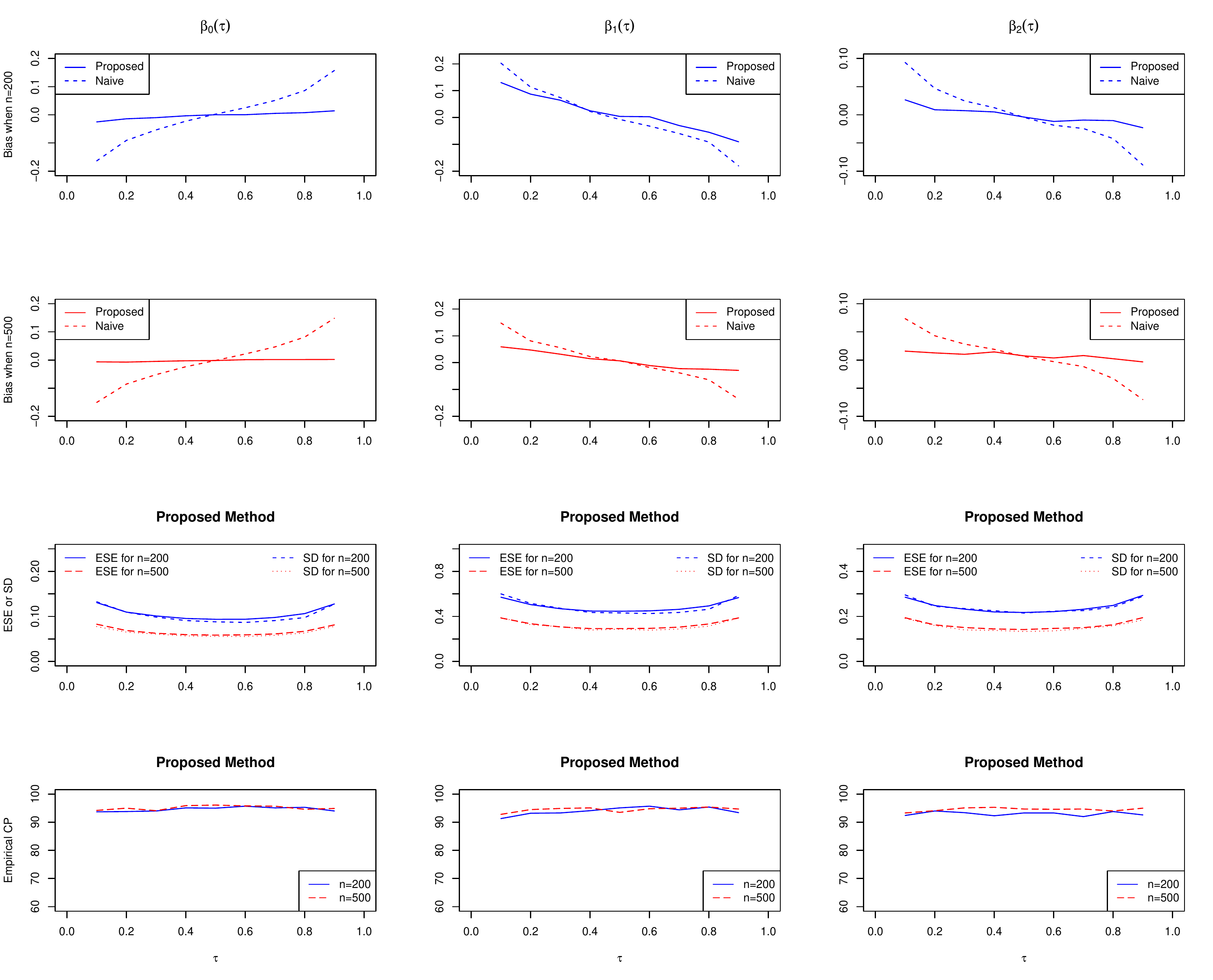}
\caption{Simulation results for Case 1. %where errors are generated from Laplace distribution and they are independent of the covariates.
Lines in blue correspond to the results with $n=200$ and lines in red correspond to the results with $n=500$. ESE stands for the estimated standard error, SD stands for the empirical standard deviation, and CP stands for the coverage probability of a 95\% confidence interval. }
\label{fig1}
\end{figure}

In Figure \ref{fig1},
 %and \ref{fig2},
we present the simulation results for {Case 1}. We relegate the results for Cases 2, 3 and 4 to the Supplementary Materials. In the figures we plot the empirical bias, the empirical standard deviation (SD), the empirical coverage probability of 95\% confidence intervals, and the average of estimated standard deviations of the proposed estimator over $\tau$. It is clear that the proposed estimator yields considerable bias reductions compared to the naive estimator. The bias reductions are most prominent at $\tau$'s close to $0$ or $1$. The bias of the proposed estimator deceases towards $0$ as the sample size increases. We also see from Figures \ref{fig1} %and \ref{fig2}
that proposed resampling based SD estimates well match the empirical SDs, and decrease with the sample size at the expected rate. The empirical coverage probabilities are close to the nominal level for all $\tau$'s and both sample sizes. We have very similar observations for the simulations in Cases 2{--}4 (see the Supplementary Materials, Figures {S1--S3}).

We further examine the robustness of our approach to the misspecification of the trajectory error distribution. Specifically, we generate the $\epsilon_{ij}$ {%\color{red} in \eqref{simumodel1}(delete because ??)}
from the uniform distribution $ Unif(-\sqrt{3}/2, \sqrt{3}/2)$, while applying $\hat\bbeta(\tau)$ that assumes the Laplace or Normal trajectory errors. The simulation results are given in Figure \ref{fig3}. We observe through comparing Figure \ref{fig3} and Figure \ref{fig1} that the estimator that does not assume the correct trajectory error distribution (Uniform versus Laplace) only yields slight elevated estimation bias. Given the sample size $n=500$, the empirical bias seems quite negligible. The impact of misspecifying the error distribution on the variance estimation and coverage probabilities is also minimal, particularly for the larger sample size.  These results demonstrate the robust performance of the proposed estimator.

\begin{figure}
\centering
\includegraphics[width=16cm, height=16cm]{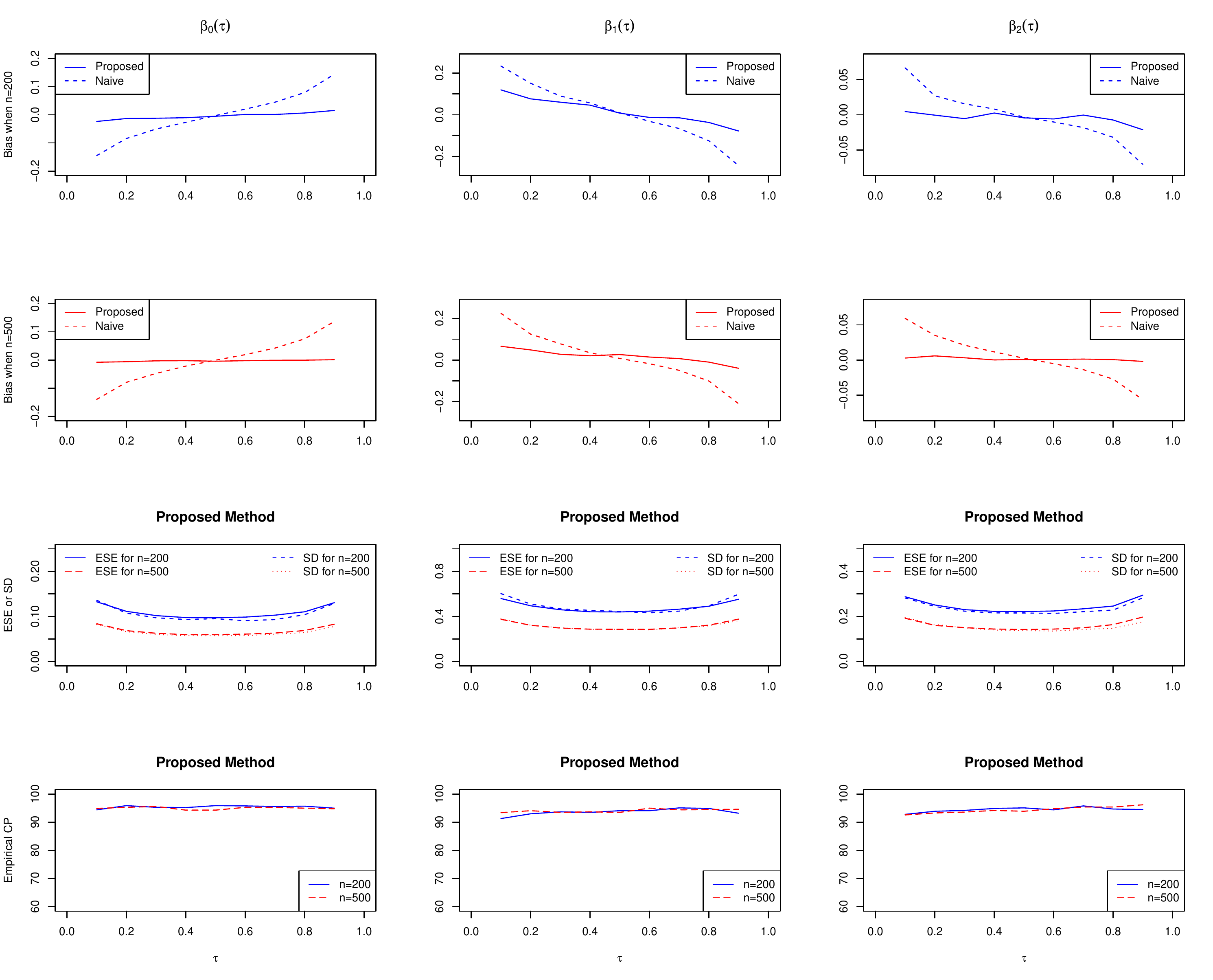}
\caption{Simulation results for the robustness study. Lines in blue correspond to the results with $n=200$ and lines in red correspond to the results with $n=500$. ESE stands for the estimated standard error, SD stands for the empirical standard deviation, and CP stands for the coverage probability of a 95\% confidence interval.}
\label{fig3}
\end{figure}

\begin{figure}
\centering
\includegraphics[width=16cm, height=16cm]{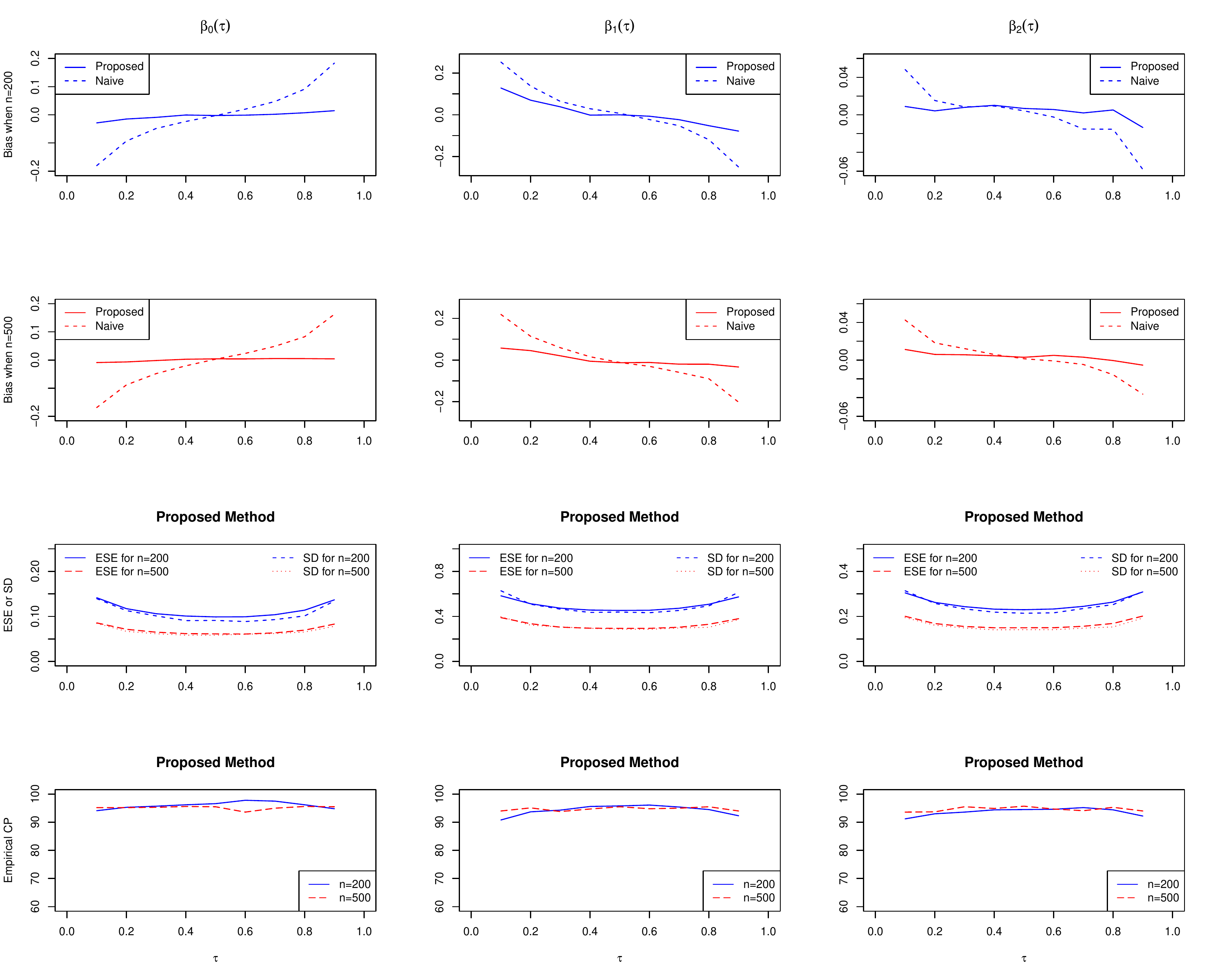}
\caption{Simulation results for quadratic trajectory model with Laplace errors. Lines in blue correspond to the results with $n=200$ and lines in red correspond to the results with $n=500$. ESE stands for the estimated standard error, SD stands for the empirical standard deviation, and CP stands for the coverage probability of a 95\% confidence interval.}
\label{fig4}
\end{figure}

In addition, we consider a setting where the longitudinal trajectory model is not linear over time, as often encountered in practice. Specifically, the longitudinal observations are generated from the following model:
\begin{eqnarray*}
Y_{ij} = a_i + b_i t_{ij}+ c_i t_{ij}^2 + \epsilon_{ij}/(1+X_{1i}), ~~~~~~~~j=1, \ldots, m_i; ~i=1, \ldots, n,
\end{eqnarray*}
and the trajectory feature is the changing rate of $Y_i(t)$ at a given time point $t_*$, i.e.
%\begin{eqnarray*}
$B_i= b_i + 2 c_i t_* = 2 + X_{1i} + X_{2i} + (0.1+X_{1i}+X_{2i})e_i$.
%\end{eqnarray*}
Here we let $a_i \sim \mbox{Exp}(0.15)$, $c_i \sim \mbox{Exp}(0.15)$,  and $t_*=1$. We generate $t_{ij}$,  $\bX_i$ and $e_i$  in the same way as before.
The results when $\epsilon_{ij}$ follows from the Laplace distribution are plotted in Figures \ref{fig4}; see the Supplementary Materials for the other results correspond to the normal error distribution. Similar to Figure \ref{fig1}, Figure \ref{fig4} suggests substantial bias reduction resulted from the proposed method as compared to the naive ones.  The coverage probabilities match quite well with the nominal value especially when $n=500$. The empirical standard errors and the estimated ones agree well and  decrease with the sample size as expected.
Overall, the performance of the proposed method are comparable between the cases with nonlinear and linear trajectory models.

\section{Analysis of the DURABLE Study}

Diabetes is a chronic disease in which there are high blood sugar levels over a prolonged period. It is a major cause of blindness, kidney failure, heart attacks, stroke and lower limb amputation and is the 7th leading cause of death affecting 422 million people worldwide \citep{roglic2016global}. Insulins are important treatments for diabetes.
The DURABLE \cite{Buse2009} trial was designed to study the efficacy, safety, and durability of two common insulin initiation regimens: twice-daily insulin lispro mixture 75/25 (LM75/25) versus once-daily insulin glargine (GL) added to oral antihyperglycemic drugs (OADs) to achieve and maintain hemoglobin A1c (HbA1c) goals.
This randomized, open label, parallel study enrolled 2187 insulin-naive patients with type 2 diabetes from 11 countries, aged 30 to 80 years, with HbA1c $> 7.0 \%$ on at least two oral antihyperglycemic agents. The HbA1c level was collected every 6 weeks during 24 weeks.

HbA1c level is an important index of glycemic control. To assess the efficacy of different treatments, we analyze the longitudinal measurements of HbA1c under the proposed quantile regression framework. Specifically,
let $Y_{ij}$ represent the $j$-th HbA1c measurements of the $i$th individual recorded in the $t_{ij}$th  week since the study enrollment ($j=1,\ldots, m_i$). After examining the observed data, we assume a quadratic trajectory model for within-subject HbA1c  measurements during the 24 week follow-up period. That is,
$$
Y_{ij} = \alpha_{i0} + \alpha_{i1} t_{ij} + \alpha_{i2} t_{ij}^2+\epsilon_{ij}.
$$
Under this model, the random intercept $\alpha_{i0}$ denotes the subject-specific baseline HbA1c measurement, and the subject-specific decreasing rate of HbA1c at a specified time point $t^*$ is given by  $-\alpha_{i1}-2\alpha_{i2}t^*$. In our analysis, we take $t^*=3$ (corresponding to 12 weeks) and set $B_i\doteq -\alpha_{i1}-2\alpha_{i2}t^*$, a meaningful trajectory feature which is not directly observed. Under this setup, we exploit how treatments and other risk factors influence the HbA1c reduction rate at 12 weeks after the initiation of the assigned insulin treatment. By adopting the quantile regression modeling, we can further delineate whether and how their associations with $B_i$ vary across diabetes patients.

The covariate of our main interest is  {\it therapy}, coded as 1 if the subject took the LM75/25 regimen and 0 if  GL regimen. We consider three other covariates according to our exploratory analysis:
{\it sulfouse}, 1 if the subject uses the sulfonylurea (SU) and 0 otherwise; {\it basfglu},  baseline fasting plasma glucose; and {\it basfins}, baseline fasting insulin. Here we standardize {\it basfglu} and {\it basfins} by subtracting their means and then dividing by their standard deviations.
Because the two regimes are added to OADs, we also include the interaction term, {\it therapy}$*${\it sulfouse}, in our model to examine if  the treatment effect (i.e. LM75/25 versus GL) on the HbA1c decreasing rate  is modified by the use of SU.
We exclude 107 subjects who have missing covariates in the analysis. We also exclude {263} subjects
%(\textcolor{yellow}{but 411 and 107 subjects have overlap})
who have less than three measurements of HbA1c, because they do not contribute information to studying {$B_{i}$}.
There are {1,717} subjects in our analysis; 71 subjects are in GL group without SU, 800 subjects are in GL group with SU, 65 subjects are in LM75/25 group without SU, and 781 subjects are in LM75/25 group with SU.  {The continuous covariate {\it basfglu} ranges from 1.04 to 26, with mean=11 and standard deviation=3.6. The {\it basfins} ranges from $-2$ to 143, with mean=10 and standard deviation=9.0.}
%Table 1 %\ref{table1} presents other summary statistics for the discrete covariates {\it therapy} and {\it sulfouse}.

%\begin{table}\caption{ { Summary statistics for the DURABLE data} }
%%\renewcommand{\arraystretch}{0.8}
%\label{table1}
%\centering
%{\footnotesize
%\fbox{
%\begin{tabular}{ccccc}
%%\hline \hline
%           &   \multicolumn{4}{c}{Category by therapy and SU use}   \\[1mm]
%    \cline{2-5} \\ [-2mm]
%           &   GL       &  LM75/25   &    GL       &  LM75/25    \\ [1mm]
%           & no SU      & no SU      &    SU       &   SU      \\
%%          &    \tabincell{c}{GL \\ no SU} & \tabincell{c}{ LM75/25 \\ no SU}  & \tabincell{c}{ GL \\ SU }  &  \tabincell{c} { LM75/25 \\ SU }  \\
%\hline
%  $n$      &   71       &    65      &    800      &   781         \\ [1mm]
% (\%)      & (4.14 \%)  &  (3.79 \%) &  (46.59 \%) &  (45.48 \%)   \\
%% \tabincell{c}{ $n$ \\ (\%) }& \tabincell{c}{71 \\ (4.14 \%)}  &  \tabincell{c}{65 \\ (3.79 \%) } &  \tabincell{c} {800 \\ (46.59 \%)}    &   \tabincell{c}{781 \\ (45.48 \%) }  \\
%%\hline
%\end{tabular}  }   }
%\end{table}

We apply the proposed method to perform quantile regression for $B_i$ on the covariates specified above at quantile levels equally spaced between $0.1$ and $0.8$ with step size $0.02$. {For selecting the smoothing parameter $h$, we employ the procedure introduced in Subsection 3.4 on a $h$-grid between $0.8$ and $1.5$ with step size 0.1.}
In Fig. \ref{fig5}, we plot the proposed estimated coefficients (red solid line) along with the 95\% pointwise confidence intervals (red dot dashed lines) for $\tau \in [0.1, 0.8]$ based on 200 resampling samples. The naive estimators (black dashed lines) are also plotted for comparisons.
In Fig. \ref{fig5}, the estimated intercept coefficients represent the estimated quantiles of HbA1c decreasing rate for subjects receiving GL therapy, no SU use, with baseline fasting glucose and baseline fasting insulin set as {10.98mmol/L and 10.06 units }(which are their observed average values in this dataset). For example, Fig. \ref{fig5} shows that the estimated median rate of HbA1c decreasing is about  {0.34} (\%) per month. Compared to the naive estimates, the proposed estimates are generally less jagged (over $\tau$) and have narrower confidence intervals. Some naive estimates even lie outside the proposed confidence intervals. These observations indicate the benefits of applying the proposed method over the naive approach.

\begin{figure}
\centering
\includegraphics[width=16cm, height=16cm]{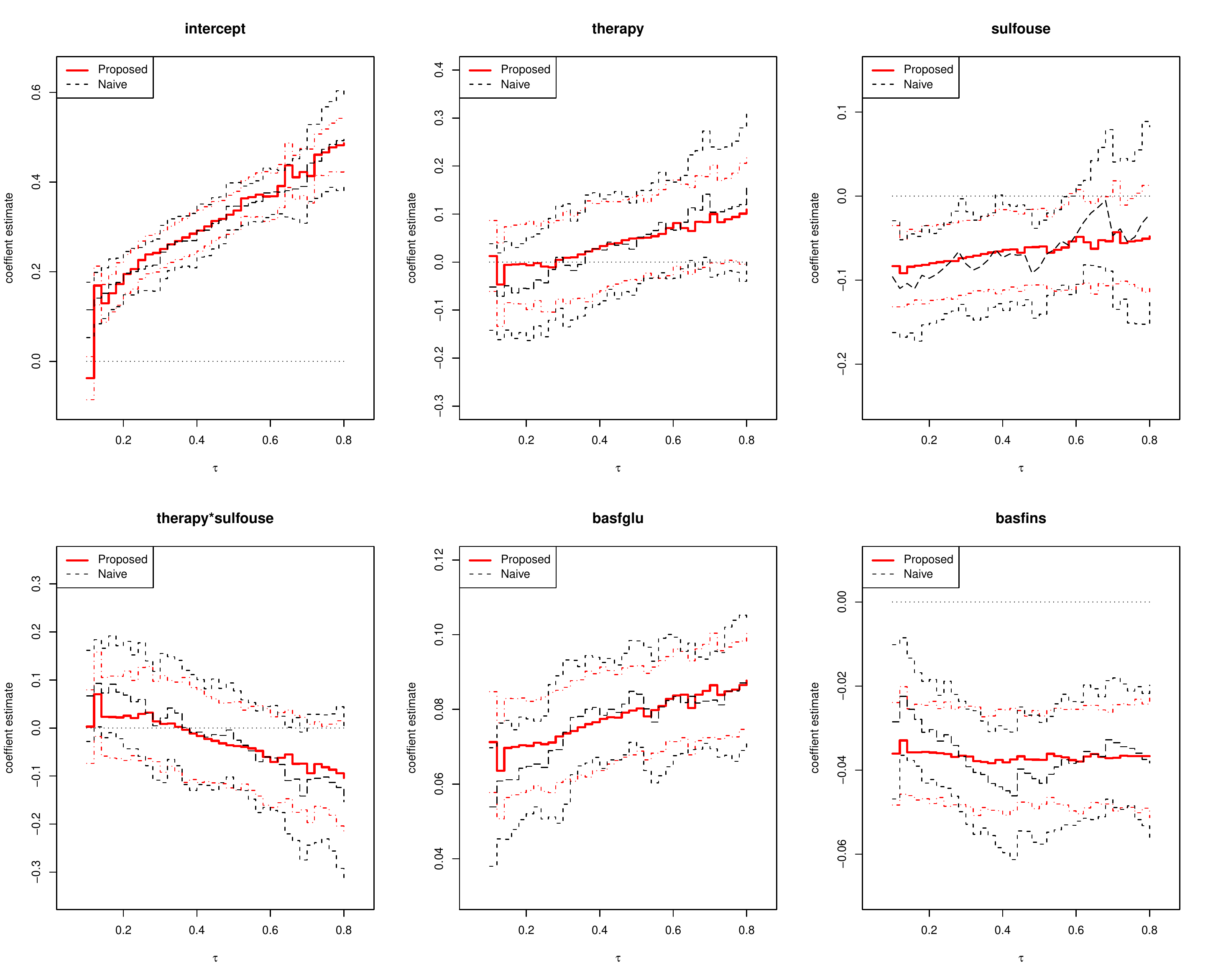}
\caption{ The DURABLE data example: the proposed coefficient estimates (red solid line) and the 95\% pointwise confidence intervals (red dot-dashed line), the naive coefficient estimates (black long-dashed line) and 95\% pointwise confidence intervals (black dashed line)   }  \label{fig5}
\end{figure}

To interpret the covariate coefficients in Figure \ref{fig5}, note that positive covariate coefficients indicate quicker HbA1c decreasing over time given one unit increase in the corresponding covariate. We observe that the coefficients for baseline fasting glucose and baseline fasting insulin are significantly different from 0 for all $\tau$'s considered. These results suggest that a higher baseline fasting glucose and a lower baseline fasting insulin are significantly associated faster decreasing of HbA1c. These are sensible findings that can explained as follows. First, a higher baseline glucose generally implicates a higher baseline HbA1c. It is reasonable to expect a more rapid HbA1c reduction for those with high baseline HbA1c values compared to those with low baseline HbA1c values. Secondly,  a lower baseline insulin may imply less insulin resistance. Consequently, it is associated with a stronger response to an insulin treatment that is manifested by a quicker HbA1c decreasing.

To evaluate the treatment effect, we examine the coefficient plots for {\it therapy}, {\it sulfuse}, and {\it therapy*sulfuse} simultaneously. In the presence of the interaction term {\it therapy*sulfuse}, the coefficients for {\it therapy} can be interpreted as the treatment effect for subjects without SU use. The estimated coefficients for {\it therapy} suggest that the therapy LM75/25 may offer some significant advantage over GL in terms of lowering HbA1c more quickly among the ``strong'' responders of the insulin treatments (corresponding to the large $\tau$'s) when SU is not used.  For the subjects with sustained HbA1c, either due to low baseline HbA1c or weak response to the insulin treatment, the GL and LM75/25 therapies demonstrate little difference in the decreasing rate of HbA1c.  The estimated coefficients for {\it sulfuse} suggest some negative impact of using SU on the treatment efficacy in lowering HbA1c. The negative impact seems diminished in ``strong'' responders of the insulin treatment.

In Fig. \ref{fig6}, we  plot the estimates for {\it therapy} coefficient $+$ {\it therapy}*{\it sulfouse} coefficient and {\it sulfouse} coefficient $+$ {\it therapy}*{\it sulfouse} coefficient, which represent the treatment effect for subjects with SU use and the effect of SU use for subjects receiving LM75/25. Comparing Fig. \ref{fig6} with the plots for {\it therapy} and {\it sulfuse} in Fig. \ref{fig5}, we see that for subjects with SU, choosing LM75/25 versus GL has little impact on the HbA1c decreasing rate, while LM75/25 demonstrates some moderate advantage over GL for subjects without SU use. This may be due to the fact that SU is a treatment for lowering postprandial blood glucose level, GL is a basal insulin to lower the overall glucose level, and LM75/25 targets both basal and postprandial. Therefore, when SU is added, both GL and LM75/25 will cover basal and postprandial blood glucose, thereby resulting in similar glycemic control, and hence similar HbA1c decreasing rates. Without SU, LM75/25 can cover postprandial glucose, while GL cannot; therefore LM75/25 demonstrates some moderate benefit over GL. % which covers basal only.
It is also shown by comparing Fig. \ref{fig6} with Fig. \ref{fig5} that for subjects treated with LM75/25, using SU may be associated with slower HbA1c reduction; its effect has a consistent direction with but a smaller magnitude than that in subjects treated with GL, particularly for $\tau$'s less than $0.5$. The overall negative impact of SU may associate with  hypoglycemic events, as SU is a known significant factor to cause hypoglycemic events, which limits titrating insulin \citep{Fu2014} and then lead to worse glycemic control. The smaller impact of SU use under LM75/25 (versus that under GL) may be explained by the bigger overlap in glucose control coverage between SU and  LM75/25.
Despite these sensible observations regarding the interaction between {\it therapy} and {\it sulfouse}, the confidence intervals for the coefficient for {\it therapy*sulfouse} (see Fig. \ref{fig5}) do not suggest statistical significance. This is likely due to the small size of the group without SU use. A larger future study may help provide a more confirmative conclusion.

\begin{figure}\label{realpred}
\centering
\includegraphics[width=16cm, height=8cm]{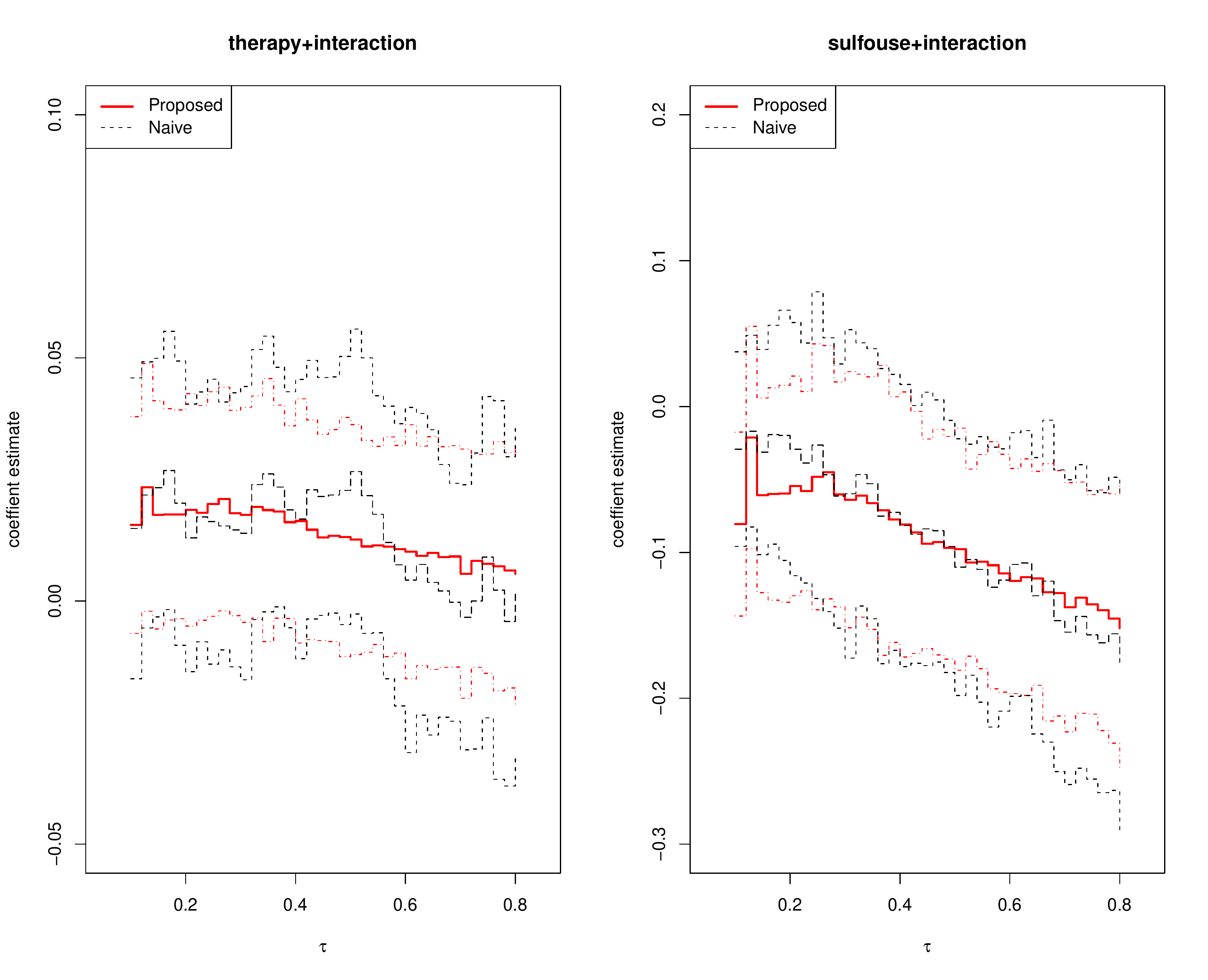}
\caption{The DURABLE data example: the estimates for {\it therapy} coefficient plus {\it therapy}*{\it sulfouse} coefficient and {\it sulfouse} coefficient plus {\it therapy}*{\it sulfouse} coefficient, and the corresponding 95\% pointwise confidence intervals. Red lines correspond to the proposed method and black lines correspond to the naive method.%(therapy=0, sulfouse=0; therapy=1, sulfouse=0; therapy=0, sulfouse=1; therapy=1, sulfouse=1)
 }
 \label{fig6}
\end{figure}

Finally, we employ the proposed test $\mathcal{T}$ to assess whether or not the effects of {\it therapy} and other covariates are constant over $\tau$. We set $\tau_L=0.1$ and $\tau_U=0.8$ and adopt $\Xi(v) = I\{v > (\tau_L+\tau_U)/2\}$. {For the coefficient for {\it therapy}, we obtain $\mathcal{T}=0.589$, which falls into the rejection region $(-\infty, -0.259) \cup (0.291, \infty)$ at level $0.05$. }This suggests that the difference between LM75/25 and GL may vary across type II diabetes patients. For example, it may differ by the status of SU use as suggested by Fig. \ref{fig5} and Fig. \ref{fig6}.
{We also reject the constancy of the effect of {\it basfglu}. This is consistent with our observation in Fig. \ref{fig5}, which is, baseline fasting glucose has a bigger influence on the upper quantiles of the HbA1c decreasing rate.} Our constancy tests suggest that the location-shift effects may be adequate for {\it sulfouse} and {\it basfins}. %This is confirmed by using $\mathcal{T}$ with a different weight function, $\Xi(v) = I\{v < (\tau_L+\tau_U)/2\}$.

It is worth mentioning that we also apply the proposed method assuming the linear trajectory model for within-subject HbA1c. We obtain quite similar results regarding the covariate effects on the decreasing rate of HbA1c at  the $12$ weeks follow-up. {The counterparts of Fig. \ref{fig5} and Fig. \ref{fig6} are presented in the Supplementary Materials; see Figures S5 and S6. }This suggests the robustness of the proposed trajectory quantile regression method to the specification of the underlying trajectory model.

\section{Discussion}

This work, to the best of our knowledge, is the first effort to utilize the device of quantile regression to explore the associates of longitudinal outcome trajectory which may follow heterogeneous patterns. As a proof of concept, we assume polynomial outcome trajectories in the present paper, while the methodology can readily be extended to accommodate nonparametric trajectories, for example, through B-Spline. The details with such extensions will be spelled out in separate work.

In this paper, we derive the regression quantiles of the latent trajectory feature when the error term in the trajectory model follows the Normal or Laplace distribution. Appealingly, they take the same practical form, as pointed out at the end of Section 3.2, and are shown to have robust performance when the trajectory errors follow a different distribution. It is worth noting that the bias correction strategy adopted in the proposed estimation procedure  can be  further extended to a wider class of distribution families, as long as their characteristic functions are proportional to the inverse of a polynomial; see  \cite{Hong2003} and \cite{Wang2012} for related discussions. This entails broader applicability and enhanced usefulness of the proposed trajectory quantile regression framework.

%%%%%%%%%%%%%%%%%%%%%%%%%%%%%%%%%%%%%%%%%%%%%%%%%%%%%%%%%%%%%%%%%%%%%%%%%%%%%%%%%%%%%%%%%%%%%%%%%%%%%%%%%%%%%%%%%%%%%%%%%%%%

%\lhead[\footnotesize\thepage\fancyplain{}\leftmark]{}\rhead[]{\fancyplain{}\rightmark\footnotesize\thepage}%Put this line in Page 2

%%%%%%%%%%%%%%%%%%%%%%%%%%%%%%%%%%%%%%%%%%%%%%%%%%%%%%%%%%%%%%%%%%%%%%%%%%%%%%%%%%%%%%%%%%%%%%%%%%%%%%%%%%%%%%%%%%%%%%%%%%%%
\vskip 14pt
\noindent {\large\bf Supplementary Materials}

Supplementary Materials, which include large sample properties, justification of the proposed resampling-based inference procedure, additional simulation studies, and linear trajectory model for DURABLE data are available online.
\par
%%%%%%%%%%%%%%%%%%%%%%%%%%%%%%%%%%%%%%%%%%%%%%%%%%%%%%%%%%%%%%%%%%%%%%%%%%%%%%%%%%%%%%%%%%%%%%%%%%%%%%%%%%%%%%%%%%%%%%%%%%%%
\vskip 14pt
\noindent {\large\bf Acknowledgements}

This work was partially supported by National Institutes of Health Grants R01HL113548.
\par

%%%%%%%%%%%%%%%%%%%%%%%%%%%%%%%%%%%%%%%%%%%%%%%%%%%%%%%%%%%%%%%%%%%%%%%%%%%%%%%%%%%%%%%%%%%%%%%%%%%%%%%%%%%%%%%%%%%%%%%%%%%%

%\pagestyle{plain}
%\def\n{\noindent}
%%\lhead[\fancyplain{} \leftmark]{}
%\chead[]{}
%%\rhead[]{\fancyplain{}\rightmark}
%\cfoot{}
%%\headrulewidth=0pt  %<-modified by Ivan
%
%%\lhead[\footnotesize\thepage\fancyplain{}\leftmark]{}\rhead[]{\fancyplain{}\rightmark\footnotesize\thepage}%Put this line in Page 2
%\lhead{}
%\rhead{\thepage}
%
%
%\markboth{\hfill{\footnotesize\rm Huijuan Ma, Limin Peng and Haoda Fu} \hfill}
%{\hfill {\footnotesize\rm Quantile Regression of Latent Longitudinal Trajectory Features} \hfill}
%
%%\iffalse
%\bibhang=1.7pc
%\bibsep=2pt
%\fontsize{9}{14pt plus.8pt minus .6pt}\selectfont
%\renewcommand\bibname{\large \bf References}
%%\begin{thebibliography}{11}
%\expandafter\ifx\csname
%natexlab\endcsname\relax\def\natexlab#1{#1}\fi
%\expandafter\ifx\csname url\endcsname\relax
%  \def\url#1{\texttt{#1}}\fi
%\expandafter\ifx\csname urlprefix\endcsname\relax\def\urlprefix{URL}\fi
%%\fi

\bibliographystyle{biom}      % Chicago style, author-year citations
\bibliography{longitudinal}   % name your BibTeX data base

%-------------------------------------------
%\vskip .65cm
%\noindent
%Huijuan Ma, Department of Biostatistics and Bioinformatics, Emory University, Atlanta, GA, 30322, U.S.A.
%\vskip 2pt
%\noindent
%E-mail: (huijuan.ma@emory.edu)
%\vskip 10pt
%
%\noindent
%Limin Peng (Corresponding author), Department of Biostatistics and Bioinformatics, Emory University, Atlanta, GA, 30322, U.S.A.
%\vskip 2pt
%\noindent
%E-mail: (lpeng@emory.edu)
%\vskip 2pt
%\noindent
%Phone: (01)404-727-7701
%\vskip 2pt
%\noindent
%Fax: (01)404-727-1370
%\vskip 10pt
%
%\noindent
%Haoda Fu, Eli Lilly and Company, Indianapolis, Indiana, U.S.A.
%\vskip 2pt
%\noindent
%E-mail: (fu$_{-}$haoda@lilly.com)
%\vskip 10pt
% \vskip .3cm
%\centerline{(Received ???? 20??; accepted ???? 20??)}\par
\end{document}

%%%%%%%%%%%%%%%%%%%%%%%%%%%%%%%%%%%%%%%%%%%%%%%%%%%%%%%%%%%%%%%%%%%%%%%%%%%%%%%%%%%%%%%%%%%%%%%%%%%%%%%%%%%%%%%%%%%%%%%%%%%%
%%%%%%%%%%%%%%%%%%%%%%%%%%%%%%%%%%%%%%%%%%%%%%%%%%%%%%%%%%%%%%%%%%%%%%%%%%%%%%%%%%%%%%%%%%%%%%%%%%%%%%%%%%%%%%%%%%%%%%%%%%%%

\renewcommand{\baselinestretch}{1}

%\markright{ \hbox{\footnotesize\rm Statistica Sinica: Supplement
%%{\footnotesize\bf 24} (201?), 000-000
%}\hfill\\[-13pt]
%\hbox{\footnotesize\rm
%%\href{http://dx.doi.org/10.5705/ss.20??.???}{doi:http://dx.doi.org/10.5705/ss.20??.???}
%}\hfill }
%
%\markboth{\hfill{\footnotesize\rm Huijuan Ma, Limin Peng and Haoda Fu} \hfill}
%{\hfill {\footnotesize\rm Quantile Regression of Latent Longitudinal Trajectory Features} \hfill}
%
%\renewcommand{\thefootnote}{}
%$\ $\par
%\fontsize{12}{14pt plus.8pt minus .6pt}\selectfont

%%%%%%%%%%%%%%%%%%%%%%%%%%%%%%%%%%%%%%%%%%%%%%%%%%%%%%%%%%%%%%%%%%%%%%%%%%%%%%%%%%%%%%%%%%%%%%%%%%%%%%%%%%%%%%%%%%%%%%%%%%%%

 \centerline{\large\bf Supplementary Materials of ``Quantile Regression of}
\vspace{2pt}
 \centerline{\large\bf Latent Longitudinal Trajectory Features"}
\vspace{.4cm}
\centerline{Huijuan Ma$^1$, Limin Peng$^1$ and Haoda Fu$^2$}
\vspace{.2cm}
\centerline{\it $^{1}$Department of Biostatistics and Bioinformatics, Emory University}
\vspace{.2cm} \centerline{\it $^{2}$Eli Lilly and Company}
\vspace{.55cm}
\fontsize{9}{11.5pt plus.8pt minus .6pt}\selectfont
%\noindent
%CONTENT OF A BRIEF NOTE.........\\
%CONTENT OF THE BRIEF NOTE.\\
%CONTENT OF THE BRIEF NOTE.\\
%\par

\setcounter{section}{0}
\setcounter{equation}{0}
\def\theequation{S\arabic{section}.\arabic{equation}}
\def\thesection{S\arabic{section}}

\fontsize{12}{14pt plus.8pt minus .6pt}\selectfont

\section{ Large Sample Properties }

{
\subsection{ Main Results }

We investigate the large sample properties of the proposed estimator. In this subsection, we establish the uniform consistency and weak convergence of the proposed estimator $\hat\bbeta_{n, h_n}(\tau)$, with a shorthand notation $\hat{\bbeta}(\tau)$ hereafter, for $\tau \in [\tau_L, \tau_U]$.

We first introduce the notation. Let $\| \cdot \|$ denote the $L_2$ norm of the corresponding vector or matrix after vectorization. Define $r_0=\inf\{ j \geq 1: \int_{-\infty}^{\infty} x^j K^{(1)}(x) d x \neq 0 \}$ and $m_0=\lim_{n \rightarrow \infty} n^{-1} \sum_{i=1}^n \int_0^{\infty} d N_i(t)$, where $N_i(t)=\sum_{j=1}^\infty I(t_{ij} \leq t)$. Let $F_B(t|\bX)=\Pr(B \leq t|\bX)$ and $f_B(t|\bX)=d F_B(t|\bX)/ d t$. Define
$ \psi_{\tau, h}^*(\mathcal{O}_i, \bb, c^2)=\partial \rho_{\tau, h}^*(\mathcal{O}_i, \bb, c^2)/\partial \bb$,
$ \psi_{\tau, h, b}^*( \mathcal{O}_i, \bb, c^2 ) = \partial \psi_{\tau, h}^*(\mathcal{O}_i, \bb, c^2) / \partial \bb$,
$ \psi_{\tau, h, c}^*( \mathcal{O}_i, \bb, c^2 ) = \partial \psi_{\tau, h}^*(\mathcal{O}_i, \bb, c^2) / \partial c^2$,
 and
 $ \bU_{\tau, h}(\bb, c^2) = n^{-1/2} \sum_{i=1}^n \psi_{\tau, h}^*(\mathcal{O}_i, \bb, c^2)$.
Further denote  $\mu_{\tau, h}(\bb, c^2)=E[ n^{-1/2} \bU_{\tau, h}(\bb, c^2) ]$, $\bB_\tau(\bb, c^2)= \lim_{n \rightarrow \infty} \partial \mu_{\tau, h}(\bb, c^2)/\partial \bb$, and $\bA_\tau(\bb, c^2) = \lim_{n \rightarrow \infty} \partial \mu_{\tau, h}(\bb, c^2)/\partial c^2$, where we suppose the sequence $h \rightarrow 0$ as $n \rightarrow \infty$.
%Note that the proposed estimators satisfy $\bU_{\tau}\{ \hat{\bbeta}(\tau), \hat{\sigma}^2\}=0$.
Let $\mu_0(\bb)=E[D^{-1/2} \bX \{ F_B( \bX^{\T}\bb |\bX )-\tau \}]$ and
$\bB_0(\bb)=E[D^{-1/2} \bX \bX^{\T} f_B( \bX^{\T}\bb |\bX )]$.

We assume the following regularity conditions:

%\begin{enumerate}
A1:  (a) The expectation $E( \| \bX_i \|^2 )$ is bounded, and $E(\bX_i \bX_i^{\T})$ is a positive definite $p \times p$ matrix; (b) $D_i=\delta(\bX_i, \bZ_i) \bgamma^{\T} \{ \int_0^{\infty} \bZ_i(t) \bZ_i(t)\trans d N_i(t) \}^{-1} \bgamma$ is bounded from infinity and is bounded away from zero, where $\bZ_i(t)$ is a piecewise constant function satisfies $\bZ_i(t_{ij})=\bZ_{ij}$, the $j$th row of $\bZ_i$; (c) $m_0$ is finite, and $\mbox{RSS}_i$
%=\int_0^{\infty} \int_0^{\infty} Y_i(s) \left[ 1- \bZ_i(s) \{ \int_0^{\infty} \bZ_i^{\T}(u) \bZ_i(u) d N_i(u) \}^{-1} \bZ_i^{\T}(t) \right] \\ Y_i(t) d N_i(s) d N_i(t)$
is bounded.

A2: (a) $K^{(j)}(\cdot)$ is uniformly bounded for $j=0,1,\ldots,4$ in the Laplace error model and for $j \geq 0$ in the normal error model; (b) $r_0 \geq 2$ and for each integer $j$ ($0 \leq j \leq r_0$), $\int_{-\infty}^{\infty} | x^j K^{(1)}(x)| d x < \infty$.

A3: For each integer $j$ such that $1 \leq j \leq r_0$, $F_B^{(j)}(t|\bX)$ is uniformly bounded over $t$ and $\bX$, where $F_B^{(j)}(t|\bX)$ is the $j$th derivative of $F_B(t|\bX)$.

A4: Each component of $\bbeta_0(\tau)$ is Lipschitz continuous for $\tau \in [\tau_L, \tau_U]$.

A5: (a) For some $d_0 > 0$ and $c_0 > 0$, $\inf_{\bb \in \mathcal{B}(d_0)} \mbox{eigmin}~ \bB_0(\bb) \geq c_0$, where $\mathcal{B}(d)=\{ \bb \in \mathrm{R}^p: \inf_{\tau \in [\tau_L, \tau_U]} \| \bb -\bbeta_0(\tau) \| \leq d_0 \}$, and $\mbox{eigmin}(\cdot)$ denotes the minimum eigenvalue of a matrix;
(b) $\inf_{\tau \in [\tau_L, \tau_U]} \mbox{eigmin} ~ \bB_{\tau}\{ \bbeta_0(\tau), \sigma^2 \} > 0$.

A6: Their exists a neighborhood of $\sigma^2$, denoted as $\mathcal{A}$, such that $\partial \mu_{\tau, h}(\bb, c^2) / \partial c^2$ is bounded uniformly in $\bbeta \in \mathcal{B}(d_0)$,
 $c^2 \in \mathcal{A}$ and $\tau \in [\tau_L, \tau_U]$ for every $h > 0$. Furthermore, the limit $\bA_\tau \{ \bbeta_0(\tau), \sigma^2 \}$ is also uniformly bounded.

A7:  (a) $\sup_{\tau \in [\tau_L, \tau_U]} \| E[ \psi_{\tau, h, b}^*\{\mathcal{O}_i, \bbeta_0(\tau), \sigma^2 \} ]^2 \|$  and
    $\sup_{\tau \in [\tau_L, \tau_U]} \| E[ \psi_{\tau, h, c}^*\{\mathcal{O}_i, \bbeta_0(\tau), \sigma^2 \} ]^2 \|$  are  bounded for every $h>0$;
       (b) $ E[ \psi_{\tau, h, b}^*\{\mathcal{O}_i, \bbeta(\tau), \sigma^2\} ]^2 $ and $E[ \psi_{\tau, h, c}^*\{\mathcal{O}_i, \bbeta(\tau), \sigma^2 \} ]^2 $ are component-wise continuous in sup-norm as a functional of $\bbeta(\tau)$ and $\sigma^2$ for every $h$.

Assumption A1 assumes the boundedness of covariates, which is often met in practice.
Assumption A2 imposes conditions on the smoothness of $K(\cdot)$. It often holds, for example, when $K(\cdot)$ is chosen as the standard normal  distribution function.
Assumption A3 gives some conditions for the conditional distribution function $F_B(t|\bX)$, which are routine in quantile regression literature.
Assumptions A4 and A5 are key to ensure the identifiability of $\bbeta_0(\tau)$, and to achieve the consistency of $\hat{\bbeta}(\tau)$ and the tightness of the limit process of $\sqrt{n}\{ \hat{\bbeta}(\tau)-\bbeta_0(\tau) \}$. Similar assumptions have been adopted in various quantile regression context, for example with censored data \citep{Peng2009} and longitudinal data \citep{Sun2016}.
By Assumption A6, the variability associated with $\hat{\sigma}^2$ has only tractable impact on the estimation of $\bbeta_0(\tau)$.
Following the arguments in \cite{Wu2015}, Assumption A7 holds for both the Laplace and the normal errors if we choose $K(\cdot)$ as the standard normal distribution function.
%Actually, Assumptions A2 and A3 are also used in censored quantile regression model with covariate measurement errors (Wu et al., 2016).

\begin{theorem} \label{theocons}
Suppose model (3) holds for $\tau \in [\tau_L, \tau_U]$. Under Assumptions A1--A6, $\sup_{\tau \in [\tau_L, \tau_U]}   \|  \hat{\bbeta}(\tau)-\bbeta_0(\tau)\| \stackrel{p}{\longrightarrow} 0 $ as $n \rightarrow \infty$.
\end{theorem}

\begin{theorem}  \label{theonorm}
Suppose model ({3}) holds for $\tau \in [\tau_L, \tau_U]$. Under Assumptions A1--A7, $n^{1/2} \{ \hat{\bbeta}(\tau) - \bbeta_0(\tau) \}$ converges weakly to a mean zero Gaussian process whose covariance at  $\tau_1, \tau_2 \in [\tau_L, \tau_U]$ is $\bB_{\tau}\{ \bbeta_0(\tau_1), \sigma^2 \}^{-1}  \bSigma(\tau_1, \tau_2) \bB_{\tau}\{ \bbeta_0(\tau_2), \sigma^2 \}^{-1}$, where
$
\bSigma(\tau_1, \tau_2) = \lim_{n \rightarrow \infty} E \{ \bxi_i(\tau_1) \bxi_i(\tau_2)^{\T} \}
$
with
$$
\bxi_i(\tau) =  \psi_{\tau, h}^* \{\mathcal{O}_i, \bbeta_0(\tau), \sigma^2 \} +  \bA_{\tau} \{ \bbeta_0(\tau), \sigma^2 \} \left\{ \frac{n}{N-qn} \mbox{RSS}_i -\sigma^2 \right\}.
$$
\end{theorem}

The proofs of Theorems 1 and 2 are provided in the { following Subsection S1.2}.

{ \subsection{Proofs of Theorems 1 and 2} }
%\section{Appendix}
%\section*{Appendix:  Proofs of Theorems 1 and 2}
\setcounter{theorem}{0}
%Before giving the proofs of theorems 1 and 2, we first give a lemma to state the large sample properties of $\hat{\sigma}^2$.
First, we present three technical lemmas that give the key results for proving Theorems 1-2.  The proofs of these lemmas are provided in the { Subsection S1.3}.

\begin{lemma} \label{lemsig}
Under Assumption A1--(c), $\hat{\sigma}^2$ converges to $\sigma^2$ almost surely when $n \rightarrow \infty$. Moreover, $\sqrt{n}(\hat{\sigma}^2 - \sigma^2)$ converges in distribution to $N(0, \zeta^2/(m_0-q)^2)$, where $\zeta^2= \mbox{Var}(\mbox{RSS}_i)$.
\end{lemma}

\begin{lemma} \label{lem5}
Under Assumptions A1--A3, $\| \mu_{\tau, h}(\bb, \sigma^2) - \mu_0(\bb) \| \leq O(h^{r_0})$ uniformly over $\bb \in \mathbb{R}^p$ as $n \rightarrow \infty$.
\end{lemma}

\begin{lemma} \label{lemlas}
For any sequence $\{ \check{\bbeta}(\tau): \tau \in [\tau_L, \tau_U] \}$ that satisfies $\sup_{\tau \in [\tau_L, \tau_U]} \|\check{\bbeta}(\tau) - \bbeta_0(\tau)\| \stackrel{p}{\longrightarrow} 0$, we have
\begin{eqnarray*}
& & \sup_{\tau \in [\tau_L, \tau_U]} \Big\|  \bU_{\tau, h} \{ \check{\bbeta}(\tau), \hat{\sigma}^2 \} - \bU_{\tau, h} \{ \bbeta_0(\tau), \sigma^2 \}   \\
        & & ~~~~~~~~~~~~ - n^{1/2} \left[ \mu_{\tau, h} \{ \check{\bbeta}(\tau), \hat{\sigma}^2 \} - \mu_{\tau, h} \{ \bbeta_0(\tau), \sigma^2 \} \right] \Big\|    \stackrel{p}{\longrightarrow} 0
\end{eqnarray*}
as $n \rightarrow \infty$.
\end{lemma}

Next, we present the proofs of Theorems 1-2.

\noindent{\bf Proof of Theorem 1: } First, for any $\bu \in \mathbb{R}^p$ satisfying $\| \bu \|=1$ and for any $\delta \geq d_0$,
\begin{eqnarray*}
\bu^{\T} \left[ \mu_0 \{ \bbeta_0(\tau) + \bu \delta \} - \mu_0 \{ \bbeta_0(\tau) \} \right]  \geq \bu^{\T} \left[ \mu_0 \{ \bbeta_0(\tau) + \bu d_0 \} - \mu_0 \{ \bbeta_0(\tau) \} \right] \geq 0.
\end{eqnarray*}
Following the arguments in the Appendix of \cite{Peng2009}, we can show that
\begin{eqnarray*}
      \| \mu_0 \{ \bbeta_0(\tau)+\bu \delta \} - \mu_0 \{ \bbeta_0(\tau) \}  \|^2 \cdot \| \bu \|^2
& \geq &  \left( \bu^{\T} \left[ \mu_0 \{ \bbeta_0(\tau) + \bu \delta \} - \mu_0 \{ \bbeta_0(\tau) \}  \right] \right)^2    \\ [2mm]
%& \geq &  \left( \bu^{\T} \left[ \mu_0 \{ \bbeta_0(\tau) + \bu d_0 \} - \mu_0 \{ \bbeta_0(\tau) \}  \right] \right)^2    \\ [2mm]
%& = &  \left(  \bu^{\T} \bB_0\{ \bbeta_0(\tau) + \bu d^*  \} \bu d_0 \right)^2  \\ [2mm]
&\geq & c_0^2 d_0^2,
\end{eqnarray*}
%where $d^* \in [0, d_0]$. Since $\bbeta_0(\tau) + \bu d^* \in \mathcal{B}(d_0)$, the last above inequality follow from
under Assumption A5--(a). Therefore, we have
$ \inf_{\bb \notin \mathcal{B}(d_0)} \| \mu_0(\bb) - \mu_0\{ \bbeta_0(\tau) \} \| \geq c_0 d_0$.

Next, by simple algebraic manipulation, we have
\begin{eqnarray*}
\mu_0\{ \hat{\bbeta}(\tau) \} - \mu_0\{ \bbeta_0(\tau) \}
 &=& n^{-1/2} \bU_{\tau, h}\{ \hat{\bbeta}(\tau), \hat{\sigma}^2 \} - \mu_0\{ \bbeta_0(\tau) \}    \\ [2mm]
 & & - \left[ n^{-1/2} \bU_{\tau, h}\{\hat{\bbeta}(\tau), \hat{\sigma}^2 \} - \mu_{\tau, h} \{ \hat{\bbeta}(\tau), \hat{\sigma}^2 \} \right]  \\[2mm]
 & & - \left[ \mu_{\tau, h}\{\hat{\bbeta}(\tau), \hat{\sigma}^2 \} - \mu_{\tau, h}\{\hat{\bbeta}(\tau), \sigma^2 \} \right]   \\ [2mm]
 & & - \left[ \mu_{\tau, h}\{\hat{\bbeta}(\tau), \sigma^2 \} - \mu_0\{ \hat{\bbeta}(\tau) \}  \right]  \\ [2mm]
 &\doteq& \mbox{I} - \mbox{II} - \mbox{III} - \mbox{IV}.
\end{eqnarray*}
By the definitions of $\hat{\bbeta}(\tau)$ and $\bbeta_0(\tau)$, $n^{-1/2} \bU_{\tau, h}\{\hat{\bbeta}(\tau), \hat{\sigma}^2\}= \bf 0$, and $\mu_0\{ \bbeta_0(\tau) \}= \bf 0$. Then $\mbox{I}=0$. For $\mbox{II}$, consider $\mathcal{G}_1=\{ \psi_{\tau, h}^*(\mathcal{O}, \bb, c^2): \bb \in \mathbb{R}^p, \tau \in [\tau_L, \tau_U] \}$ for any fixed $h>0$ and $c^2>0$. The functional class $\mathcal{G}_1$ is Glivenko-Cantelli \citep{van1996}  under Assumptions A1, A2--(a) and A4. It then follows from the Glivenko--Cantelli theorem that
$$
 \sup_{\tau \in [\tau_L, \tau_U]} \| n^{-1/2} \bU_{\tau, h}\{\hat{\bbeta}(\tau), \hat{\sigma}^2 \} - \mu_{\tau, h} \{ \hat{\bbeta}(\tau), \hat{\sigma}^2 \} \| = o(1)~~ \mbox{a.s.}
$$
for any $h > 0$.
For $\mbox{III}$, a Taylor expansion of $\mu_{\tau, h}\{\hat{\bbeta}(\tau), c^2\}$ around $\sigma^2$ implies that
\begin{eqnarray*}
 \mu_{\tau, h}\{\hat{\bbeta}(\tau), \hat{\sigma}^2\} - \mu_{\tau, h}\{\hat{\bbeta}(\tau), \sigma^2\}
 = \frac{\partial \mu_{\tau, h}\{\hat{\bbeta}(\tau), c^2\} }{ \partial c^2 } \Bigg|_{c^2=\sigma_{**}^2} \cdot ( \hat{\sigma}^2 - \sigma^2 ),
\end{eqnarray*}
where $\sigma_{**}^2$ lies between $\hat{\sigma}^2$ and $\sigma^2$.
By the uniformly bounded condition of $\partial \mu_{\tau, h}\{\bb, c^2\}/ \partial c^2 $ stated in Assumption A6 and the strong consistency of $\hat{\sigma}^2$ given in Lemma \ref{lemsig}, we have
$$
\sup_{\tau \in [\tau_L, \tau_U]} \|  \mu_{\tau, h}\{\hat{\bbeta}(\tau), \hat{\sigma}^2\} - \mu_{\tau, h}\{\hat{\bbeta}(\tau), \sigma^2\} \|= o(1)~~ \mbox{a.s.}
$$
for every $h>0$.
For $\mbox{IV}$,  Lemma \ref{lem5} implies that
$$
\sup_{\tau \in [\tau_L, \tau_U]} \| \mu_{\tau, h}\{\hat{\bbeta}(\tau), \sigma^2\} - \mu_0\{ \hat{\bbeta}(\tau)\} \| < c_0 d_0/8
$$
when $h$ is sufficiently small.
Hence, we have show that
\begin{eqnarray} \label{aaeq1}
\sup_{\tau \in [\tau_L, \tau_U]} \|   \mu_0\{ \hat{\bbeta}(\tau) \} - \mu_0\{ \bbeta_0(\tau) \} \| = o_p(1)
\end{eqnarray}
when $n$ is large enough and consequently $h$ is small.
Thus, their exists $n_0 > 0$, such that for $n \geq n_0$, $\sup_{\tau \in [\tau_L, \tau_U]} \| \mu_0\{ \hat{\bbeta}(\tau) \} - \mu_0\{ \bbeta_0(\tau) \} \| < c_0 d_0/2 $ with probability 1.
This implies that $\{ \hat{\bbeta}(\tau): \tau \in [\tau_L, \tau_U] \} \subseteq \mathcal{B}(d_0)$ with probability 1 when $n \geq n_0$. Note that
$$
\sup_{\tau \in [\tau_L, \tau_U]} \| \hat{\bbeta}(\tau) - \bbeta_0(\tau) \| = \sup_{\tau \in [\tau_L, \tau_U]} \| \bB_0\{ \breve{\bbeta}(\tau) \}^{-1} [\mu_0\{ \hat{\bbeta}(\tau) \} - \mu_0\{ \bbeta_0(\tau) \} ]  \|,
$$
where $\breve{\bbeta}(\tau)$ is between $\hat{\bbeta}(\tau)$ and $\bbeta_0(\tau)$ and lies in $\mathcal{B}(d_0)$ when $n \geq n_0$.
Following (\ref{aaeq1}) and Assumption A5, we immediately get the uniform consistency of $\hat{\bbeta}(\tau)$.
\\

\noindent{\bf Proof of Theorem 2:} According to Lemma \ref{lemlas} and $\bU_{\tau, h} \{ \hat{\bbeta}(\tau), \hat{\sigma}^2 \} = 0$, we have
\begin{eqnarray*}
 & & - \bU_{\tau, h} \{ \bbeta_0(\tau), \sigma^2 \}   \\ [2mm]
 &=& n^{1/2} \left\{ \mu_{\tau, h} \{ \hat{\bbeta}(\tau), \hat{\sigma}^2 \} - \mu_{\tau, h} \{ \bbeta_0(\tau), \sigma^2 \}  \right\} + o_{[\tau_L, \tau_U]}(1)  \\ [2mm]
 &=&   \bB_{\tau}\{ \bbeta_0(\tau), \sigma^2 \}   \cdot n^{1/2} \{ \hat{\bbeta}(\tau)-\bbeta_0(\tau) \}  \\[2mm]
    && + \bA_{\tau} \{ \bbeta_0(\tau), \sigma^2 \} \cdot n^{1/2} ( \hat{\sigma}^2 - \sigma^2 ) + o_{[\tau_L, \tau_U]}(1),
\end{eqnarray*}
where $o_{[\tau_L, \tau_U]}(1)$ denotes a term that converges uniformly to 0 in probability in $\tau \in [\tau_L, \tau_U]$.
Hence, under Assumption A5--(b),
\begin{eqnarray}
& & n^{1/2} \{ \hat{\bbeta}(\tau) - \bbeta_0(\tau) \} \nonumber\\ [2mm]
&=& - \bB_{\tau}\{ \bbeta_0(\tau), \sigma^2 \}^{-1} \left[ \bU_{\tau, h} \{ \bbeta_0(\tau), \sigma^2 \}
         +  \bA_{\tau} \{ \bbeta_0(\tau), \sigma^2 \} \cdot n^{1/2} ( \hat{\sigma}^2 - \sigma^2 ) \right]   \nonumber \\ [2mm]
& & ~~~~~~~~~~~~~~~~~~~~~~~~~~~~~~ + o_{[\tau_L, \tau_U]}(1) \nonumber \\ [2mm]
&=& - \bB_{\tau}\{ \bbeta_0(\tau), \sigma^2 \}^{-1} \Bigg[ n^{-1/2} \sum_{i=1}^n \psi_{\tau, h}^* \{\mathcal{O}_i, \bbeta_0(\tau), \sigma^2 \}   \nonumber\\ [2mm]
      &&~~ + n^{-1/2} \sum_{i=1}^n  \bA_{\tau} \{ \bbeta_0(\tau), \sigma^2 \} \left( \frac{n}{N-qn} \mbox{RSS}_i - \sigma^2 \right)   \Bigg] + o_{[\tau_L, \tau_U]}(1).   \label{eqs12}
\end{eqnarray}
For any $h>0$, consider $\mathcal{G}_2=\{ \psi_{\tau, h}^*\{\mathcal{O}_i, \bbeta_0(\tau), \sigma^2\} + \bA_{\tau} \{ \bbeta_0(\tau), \sigma^2 \}  ( \frac{n}{N-qn} \mbox{RSS}_i - \sigma^2 ): \tau \in [\tau_L, \tau_U] \}$. Under Assumptions A1, A2--(a), A4 and A6, $\mathcal{G}_2$ is a uniformly bounded class, the entropy number of which behaves polynomially in $1/\epsilon$ for every $\epsilon > 0$ \citep{van1996}.
Note that $h$ depends on $n$ and goes to $0$ as $n \rightarrow \infty$.
According to Theorem 2.11.22 in \cite{van1996} that states the central limit theorem for classes of functions changing with $n$,
$$
n^{-1/2} \sum_{i=1}^n \left\{ \psi_{\tau, h}^* \{\mathcal{O}_i, \bbeta_0(\tau), \sigma^2 \} + \bA_{\tau} \{ \bbeta_0(\tau), \sigma^2 \} \left( \frac{n}{N-qn} \mbox{RSS}_i - \sigma^2 \right)  \right\}
$$
converges weakly to a zero-mean Gaussian process for $\tau_1, \tau_2 \in [\tau_L, \tau_U]$ with covariance matrix
$
\bSigma(\tau_1, \tau_2) = \lim_{n \rightarrow \infty} E \{ \bxi_i(\tau_1) \bxi_i(\tau_2)^{\T} \},
$
where
$$
\bxi_i(\tau) =  \psi_{\tau, h}^* \{\mathcal{O}_i, \bbeta_0(\tau), \sigma^2 \} +  \bA_{\tau} \{ \bbeta_0(\tau), \sigma^2 \} \left\{ \frac{n}{N-qn} \mbox{RSS}_i -\sigma^2 \right\}.
$$
It then follows from (\ref{eqs12}) that $n^{1/2} \{ \hat{\bbeta}(\tau) - \bbeta_0(\tau) \}$ converges weakly to a Gaussian process with covariance matrix
$
\bB_{\tau}\{ \bbeta_0(\tau_1), \sigma^2 \}^{-1} \bSigma(\tau_1, \tau_2) \bB_{\tau}\{ \bbeta_0(\tau_2), \sigma^2 \}^{-1}.
$
}

\subsection{ { Proofs of Technical Lemmas} }

%\begin{lemma} \label{lemsig}
%Under Assumption A1--(c), $\hat{\sigma}^2$ converges to $\sigma^2$ almost surely when $n \rightarrow \infty$. Moreover, $\sqrt{n}(\hat{\sigma}^2 - \sigma^2)$ converges in distribution to $N(0, \zeta^2/(m_0-q)^2)$, where $\zeta^2= Var(\mbox{RSS}_i)$.
%\end{lemma}
\noindent{\bf Proof of Lemma \ref{lemsig}:}
First, given the observation times $m_i=\int_0^\infty d N_i(t)$, we can show that $E(\mbox{RSS}_i) = ( m_i -q) \sigma^2$. This is because
\begin{eqnarray*}
E(\mbox{RSS}_i) &=& E(\tilde{\bY}_i^{\T} (I_{m_i}-\bP_i) \tilde{\bY}_i)= E(\tilde{\bepsilon}_i^{\T} (I_{m_i}-\bP_i) \tilde{\bepsilon}_i) \\ [2mm]
                &=& \sigma^2 \cdot \mbox{trace}(I_{m_i}-\bP_i) = (m_i-q) \sigma^2.
\end{eqnarray*}
Under Assumption A1--(c),  it follows from the strong law of large numbers for triangular arrays that $\hat{\sigma}^2$ converges almost surely to $\sigma^2$.

Moreover, by the Lindeberg--Feller central limit theorem for triangular arrays, we have
\begin{eqnarray*}
\frac{n^{-1/2} \sum_{i=1}^n ( \frac{n}{N-qn} \mbox{RSS}_i -\sigma^2 ) }{ \frac{1}{n} \sum_{i=1}^n ( \frac{n}{N-qn} )^2 \mbox{Var}( \mbox{RSS}_i )  }
= \frac{ \sqrt{n} ( \hat{\sigma}^2 -\sigma^2 ) }{ \frac{1}{n} \sum_{i=1}^n ( \frac{n}{N-qn} )^2 \mbox{Var}( \mbox{RSS}_i )  }
\end{eqnarray*}
converges in distribution to $N(0, 1)$. Note that $\frac{1}{n} \sum_{i=1}^n ( \frac{n}{N-qn} )^2 \mbox{Var}( \mbox{RSS}_i )$ converges to  $\mbox{Var}( \mbox{RSS}_i )/(m_0-q)^2$.
It then follows from the Slutsky's theorem that $\sqrt{n}(\hat{\sigma}^2 - \sigma^2)$ converges in distribution to $N(0, \zeta^2/(m_0-q)^2)$.

%\begin{lemma} \label{lem5}
%Under Assumptions A1--A3, $\| \mu_{\tau, h}(\bb, \sigma^2) - \mu_0(\bb) \| \leq O(h^{r_0})$ uniformly over $\bb \in \mathbb{R}^p$ as $n \rightarrow \infty$.
%\end{lemma}
\noindent{\bf Proof of Lemma \ref{lem5}:} Note that
\begin{eqnarray*}
 \| \mu_{\tau, h}(\bb, \sigma^2) - \mu_0(\bb) \|  & = &  \| E[\psi_{\tau, h}^*(\mathcal{O}_i, \bb, \sigma^2)] - \mu_0(\bb) \|  \\ [2mm]
& = & \left\|  \frac{\partial}{\partial \bb} E[ \rho_{\tau, h}^*(\mathcal{O}_i, \bb, \sigma^2) | \mathcal{U}_i ] - \mu_0(\bb)  \right\|   \\ [2mm]
& = & \left\| \frac{\partial}{\partial \bb} E[\rho_{\tau, h}(\mathcal{U}_i, \bb)] - \mu_0(\bb)  \right\|,
\end{eqnarray*}
where the first term
\begin{eqnarray*}
& & \frac{\partial}{\partial \bb} E[\rho_{\tau, h}(\mathcal{U}_i, \bb)]  \\ [2mm]
& = & \frac{\partial}{\partial \bb} E \left[ D_i^{-1/2} (B_i-\bX_i^{\T}\bb) \left\{ \tau - 1 + K\left( \frac{D_i^{-1/2} (B_i-\bX_i^{\T}\bb) }{h} \right) \right\} \right]  \\[2mm]
& = & E \left[ - D_i^{-1/2} \bX_i \left\{ \tau - 1 + K\left( \frac{D_i^{-1/2} (B_i-\bX_i^{\T}\bb) }{h} \right) \right\} \right]  \\[2mm]
     && + E \left[ D_i^{-1/2} (B_i-\bX_i^{\T}\bb) K^{(1)} \left( \frac{D_i^{-1/2} (B_i-\bX_i^{\T}\bb) }{h} \right) \frac{- D_i^{-1/2} \bX_i }{h} \right],
\end{eqnarray*}
and the second term
$$
\mu_0(\bb) = E \left[ D^{-1/2} \bX [ I\{ D^{-1/2}(B-\bX^{\T}\bb) < 0 \}  - \tau ] \right]
$$
provided $D^{-1/2}$ is positive definite as implied by A1--(b).
Thus
\begin{eqnarray}
& &  \| \mu_{\tau, h}(\bb, \sigma^2) - \mu_0(\bb) \| \nonumber \\[2mm]
& \leq & \left\| E \left[ D_i^{-1/2} \bX_i \left\{  K\left( \frac{D_i^{-1/2} (B_i-\bX_i^{\T}\bb) }{h} \right) - I\{D_i^{-1/2}(B_i-\bX_i^{\T}\bb) > 0 \} \right\} \right] \right\| \nonumber \\[2mm]
& & + \left\| E \left[ D_i^{-1/2} \bX_i  \frac{ D_i^{-1/2} (B_i-\bX_i^{\T}\bb)  }{h}  K^{(1)} \left( \frac{D_i^{-1/2} (B_i-\bX_i^{\T}\bb) }{h} \right)  \right] \right\| \nonumber  \\ [2mm]
& \doteq & \| I_{h1}(\bb) \| + \| I_{h2}(\bb) \|.
\label{lemma2_1}
\end{eqnarray}
Write
\begin{eqnarray}
I_{h2}(\bb) &=& E \left[ D^{-1/2} \bX ~ E \left\{ \frac{ D^{-1/2} (B-\bX^{\T}\bb)  }{h}  K^{(1)} \left( \frac{D^{-1/2} (B-\bX^{\T}\bb) }{h} \right) \Bigg| \bZ, \bX \right\}  \right]  \nonumber \\[2mm]
         &=& E \left[ D^{-1/2} \bX \int_{-\infty}^{\infty} t K^{(1)}(t) f_B(D^{1/2}ht+\bX^{\T}\bb|\bX)  D^{1/2}h dt \right]  \nonumber \\ [2mm]
         &=& E \left[ \bX \int_{-\infty}^{\infty}  K^{(1)}(t) ht f_B(D^{1/2}ht+\bX^{\T}\bb|\bX)   dt \right].   \label{aaeq2}
\end{eqnarray}
A Taylor series expansion implies that
\begin{eqnarray}
& & f_B( D^{1/2}ht+\bX^{\T}\bb |\bX ) \nonumber \\ [2mm]
&=& \sum_{i=0}^{r_0-2} \frac{1}{i!} f_B^{(i)}(\bX^{\T}\bb|\bX)(D^{1/2}ht)^i + \frac{1}{(r_0-1)!} f_B^{(r_0-1)}(t^*|\bX) (D^{1/2}ht)^{r_0-1},  \label{aaeq3}
\end{eqnarray}
where $t^*$ is between $\bX^{\T}\bb$ and $D^{1/2}ht+\bX^{\T}\bb$. By substituting (\ref{aaeq3}) into (\ref{aaeq2}), we can obtain
$$
I_{h2}(\bb) = E \left[ D^{(r_0-1)/2}\bX \frac{h^{r_0}}{(r_0-1)!}  \int_{-\infty}^{\infty} t^{r_0} K^{(1)}(t)  f_B^{(r_0-1)}(t^*|\bX)   dt \right].
$$
Using Assumptions A1--(a), A1--(b), A2--(b), A3, and the Lebesgue dominated convergence theorem, we have that
$$
h^{-r_0} I_{h2}(\bb) \rightarrow \int_{-\infty}^{\infty} t^{r_0} K^{(1)}(t) d t~\frac{1}{(r_0-1)!}  E \left[ D^{(r_0-1)/2} \bX f_B^{(r_0-1)}(\bX^{\T}\bb|\bX) \right].
$$
Hence
\begin{equation}
\sup_{\bb \in \mathbb{R}^p} \| I_{h2}(\bb) \| = O(h^{r_0})       \label{eqs16}
\end{equation}
under Assumptions A1--A3.

Next, we show that $\sup_{\bb \in \mathbb{R}^p} \| I_{h1}(\bb) \|=O(h^{r_0})$. Using the Taylor series expansion and integration by parts, we have
\begin{eqnarray*}
& & I_{h1}(\bb)   \\ [2mm]
&=& E \left[ D^{-1/2} \bX ~E \left\{  K\left( \frac{D^{-1/2} (B-\bX^{\T}\bb) }{h} \right) - I\{D^{-1/2}(B-\bX^{\T}\bb) > 0 \} \Bigg| \bZ, \bX \right\} \right]   \\[2mm]
         &=& E \left[ D^{-1/2} \bX  \int_{-\infty}^{\infty} \left\{  K\left( \frac{D^{-1/2} (t-\bX^{\T}\bb) }{h} \right) - I \{ D^{-1/2}(t-\bX^{\T}\bb) > 0 \} \right\} d F_B(t|\bX)  \right]   \\[2mm]
         &=& E \left[ D^{-1/2} \bX  \int_{-\infty}^{\infty} \left\{  K\left( t \right) - I(t > 0 ) \right\} d F_B( D^{1/2} ht+\bX^{\T} \bb|\bX)  \right]   \\[2mm]
         &=& E \left[ D^{-1/2} \bX   \left\{ F_B( \bX^{\T}\bb |\bX) - \int_{-\infty}^{\infty} K^{(1)}(t) F_B( D^{1/2}ht+\bX^{\T}\bb |\bX)   d t \right\}   \right]   \\[2mm]
         &=& - E \left[ D^{-1/2} \bX \int_{-\infty}^{\infty} \frac{1}{r_0!} F_B^{(r_0)}(t^\dag|\bX) (D^{1/2}ht)^{r_0} K^{(1)}(t) d t   \right],
\end{eqnarray*}
where $t^{\dag}$ is between $\bX^{\T} \bb$  and $D^{1/2} ht+\bX^{\T} \bb$.
Based on similar arguments  for $I_{h2}(\bb)$, we can also obtain that
\begin{equation}
\sup_{\bb \in \mathbb{R}^p } \| I_{h1}(\bb) \| = O(h^{r_0}).     \label{eqs17}
\end{equation}
By (\ref{lemma2_1}), (\ref{eqs16}), and (\ref{eqs17}), we get
$$
\sup_{\bb \in  \mathbb{R}^p } \| \mu_\tau(\bb, \sigma^2) -\mu_0(\bb) \| \leq O(h^{r_0}).
$$

%\begin{lemma} \label{lemlas}
%For any sequence $\{ \check{\bbeta}(\tau): \tau \in [\tau_L, \tau_U] \}$ that satisfies $\sup_{\tau \in [\tau_L, \tau_U]} \|\check{\bbeta}(\tau) - \bbeta_0(\tau)\| \stackrel{p}{\longrightarrow} 0$, we have
%\begin{eqnarray*}
%& & \sup_{\tau \in [\tau_L, \tau_U]} \Big\|  \bU_{\tau, h} \{ \check{\bbeta}(\tau), \hat{\sigma}^2 \} - \bU_{\tau, h} \{ \bbeta_0(\tau), \sigma^2 \}   \\
%& &~~~~~~~~~~~~~~~~~~~~~- n^{1/2} \left[ \mu_{\tau, h} \{ \check{\bbeta}(\tau), \hat{\sigma}^2 \} - \mu_{\tau, h} \{ \bbeta_0(\tau), \sigma^2 \} \right] \Big\| \stackrel{p}{\longrightarrow} 0
%\end{eqnarray*}
%as $n \rightarrow \infty$.
%\end{lemma}

\noindent{\bf Proof of Lemma \ref{lemlas}:}
If we can show
\begin{eqnarray*}
\sup_{\tau \in [\tau_L, \tau_U]} \mbox{Var} \left[ \psi_{\tau, h}^*\{ \check{\bbeta}(\tau), \hat{\sigma}^2 \} - \psi_{\tau, h}^*\{ \bbeta_0(\tau), \sigma^2 \}  \right] \stackrel{p}{\longrightarrow} 0,
\end{eqnarray*}
then Lemma \ref{lemlas} follows by using the results in \cite{Alexander1984} and the arguments for theorem 1 of \cite{Lai1988}.
Taylor series expansion and the equivalence of the finite-dimensional norms of the matrix imply that
\begin{eqnarray*}
  &  &  \mbox{Var} \left[ \psi_{\tau, h}^*\{ \check{\bbeta}(\tau), \hat{\sigma}^2 \} - \psi_{\tau, h}^*\{ \bbeta_0(\tau), \sigma^2 \}  \right]   \\ [2mm]
& \leq & \| E [ \psi_{\tau, h, b}^*\{ \bbeta_*(\tau), \sigma_*^2 \} ]^2 \| \cdot \| \check{\bbeta}(\tau) - \bbeta_0(\tau) \|^2   \\ [2mm]
& &       + \| E[\psi_{\tau, h, c}^*\{ \bbeta_*(\tau), \sigma_*^2 \}]^2 \| \cdot \| \hat{\sigma}^2 - \sigma^2 \|^2,
\end{eqnarray*}
where $\bbeta_*(\tau)$ is between $\check{\bbeta}(\tau)$ and $\bbeta_0(\tau)$, $\sigma_*^2$ is between $\hat{\sigma}^2$ and $\sigma^2$.
For every $h > 0$, under Assumption A7--(a), we can find some $C_0 > 0$ and $n_1$ such that for $n > n_1$,
$$
\sup_{\tau \in [\tau_L, \tau_U]} \| E[ \psi_{\tau, h, b}^*\{ \bbeta_0(\tau), \sigma^2 \} ]^2 \|  \leq C_0^2/4
$$
and
$$
\sup_{\tau \in [\tau_L, \tau_U]} \| E[ \psi_{\tau, h, c}^*\{ \bbeta_0(\tau), \sigma^2 \} ]^2 \|  \leq C_0^2/4.
$$
Under Assumption A7--(b), we have
$$
\sup_{\tau \in [\tau_L, \tau_U]} \| E [ \psi_{\tau, h, b}^*\{ \bbeta_*(\tau), \sigma_*^2 \} ]^2 -E [ \psi_{\tau, h, b}^*\{ \bbeta_0(\tau), \sigma^2 \} ]^2   \| \leq C_0^2/4.
$$
and
$$
\sup_{\tau \in [\tau_L, \tau_U]} \| E [ \psi_{\tau, h, c}^*\{ \bbeta_*(\tau), \sigma_*^2 \} ]^2 -E [ \psi_{\tau, h, c}^*\{ \bbeta_0(\tau), \sigma^2 \} ]^2  \| \leq C_0^2/4.
$$
Furthermore, notice that $\sup_{\tau \in [\tau_L, \tau_U]} \| \check{\bbeta}(\tau) -\bbeta_0(\tau) \|  \stackrel{p}{\longrightarrow} 0$, and  $\hat{\sigma}^2  \stackrel{p}{\longrightarrow} \sigma^2$ by Lemma \ref{lemsig}.
 Hence, for any $\epsilon > 0$ and $\eta > 0$, there exists $n_2$ such that for $n > n_2$,
$$
\Pr \left( \sup_{\tau \in [\tau_L, \tau_U]} \| \check{\bbeta}(\tau)-\bbeta_0(\tau) \| > \sqrt{\eta}/C_0  \right) < \epsilon/2
$$
and
$$
\Pr\left( \| \hat{\sigma}^2 - \sigma^2 \| > \sqrt{\eta}/C_0 \right) < \epsilon/2.
$$
Consider the case $\sup_{\tau \in [\tau_L, \tau_U]} \| \check{\bbeta}(\tau)-\bbeta_0(\tau) \| \leq \sqrt{\eta}/C_0$ and $\| \hat{\sigma}^2 - \sigma^2 \| \leq \sqrt{\eta}/C_0$.
It follows immediately that
$$
\sup_{\tau \in [\tau_L, \tau_U]}  \mbox{Var} \left[ \psi_{\tau, h}^*\{ \check{\bbeta}(\tau), \hat{\sigma}^2 \} - \psi_{\tau, h}^*\{ \bbeta_0(\tau), \sigma^2 \}  \right] \leq \eta.
$$
Therefore, for any $\epsilon > 0$, $\eta > 0$, and $n > n_3=\max(n_1, n_2)$,
\begin{eqnarray*}
 & & \Pr \left( \sup_{\tau \in [\tau_L, \tau_U]} \mbox{Var} \left[ \psi_{\tau, h}^*\{ \check{\bbeta}(\tau), \hat{\sigma}^2 \} - \psi_{\tau, h}^*\{ \bbeta_0(\tau), \sigma^2 \}  \right] > \eta  \right)   \\ [2mm]
 & \leq & \Pr \left( \| \check{\bbeta}(\tau)-\bbeta_0(\tau) \| > \sqrt{\eta}/C_0 \right) + \Pr \left( \| \hat{\sigma}^2 - \sigma^2 \| > \sqrt{\eta}/C_0  \right) \\[2mm]
 & < & \epsilon.
\end{eqnarray*}
This implies that
\begin{eqnarray*}
\sup_{\tau \in [\tau_L, \tau_U]} \mbox{Var} \left[ \psi_{\tau, h}^*\{ \check{\bbeta}(\tau), \hat{\sigma}^2 \} - \psi_{\tau, h}^*\{ \bbeta_0(\tau), \sigma^2 \}  \right] \stackrel{p}{\longrightarrow} 0.
\end{eqnarray*}

\section{Justification of the Proposed Resampling-based Inference Procedure}
Given $E(\omega_i)=1$, following the arguments in the proof of Theorem 2, we can obtain
\begin{eqnarray*}
& & n^{1/2} \{ \bbeta^*(\tau) - \hat{\bbeta}(\tau) \} \\ [2mm]
&=& - \bB_{\tau}\{ \bbeta_0(\tau), \sigma^2 \}^{-1} \Bigg[ n^{-1/2} \sum_{i=1}^n \omega_i~ \psi_{\tau, h}^* \{\mathcal{O}_i, \hat{\bbeta}(\tau), \hat{\sigma}^2 \}   \\ [2mm]
      &&~~ + n^{-1/2} \sum_{i=1}^n  \omega_i~ \bA_{\tau} \{ \bbeta_0(\tau), \sigma^2 \} \left( \frac{n}{N-qn} \mbox{RSS}_i - \sigma^2 \right)   \Bigg] + o_{[\tau_L, \tau_U]}(1).
\end{eqnarray*}
Because
$$
 n^{-1/2} \sum_{i=1}^n \psi_{\tau, h}^* \{\mathcal{O}_i, \hat{\bbeta}(\tau), \hat{\sigma}^2 \}  = o_{[\tau_L, \tau_U]}(1),
$$
and
$$
 n^{-1/2} \sum_{i=1}^n  \left( \frac{n}{N-qn} \mbox{RSS}_i - \hat{\sigma}^2 \right)  = o_{[\tau_L, \tau_U]}(1),
$$
we have
\begin{eqnarray*}
& & n^{1/2} \{ \bbeta^*(\tau) - \hat{\bbeta}(\tau) \} \\ [2mm]
&=& - \bB_{\tau}\{ \bbeta_0(\tau), \sigma^2 \}^{-1} \Bigg[ n^{-1/2} \sum_{i=1}^n (1-\omega_i) \psi_{\tau, h}^* \{\mathcal{O}_i, \bbeta_0(\tau), \sigma^2 \}   \\ [2mm]
      &&~~ + n^{-1/2} \sum_{i=1}^n (1-\omega_i) \bA_{\tau} \{ \bbeta_0(\tau), \sigma^2 \} \left( \frac{n}{N-qn} \mbox{RSS}_i - \sigma^2 \right)   \Bigg] + o_{[\tau_L, \tau_U]}(1)  \\ [2mm]
&=& - \bB_{\tau}\{ \bbeta_0(\tau), \sigma^2 \}^{-1} \cdot n^{-1/2} \sum_{i=1}^n (1-\omega_i) \bxi_i(\tau)  + o_{[\tau_L, \tau_U]}(1).
\end{eqnarray*}
Using the fact that $\mbox{var}(\omega_i)=1$, by the arguments of \cite{Lin1993}, the conditional distribution of $n^{1/2} \{ \bbeta^*(\tau) - \hat{\bbeta}(\tau) \}$ given the observed data is asymptotically equivalent to the unconditional distribution of $n^{1/2} \{ \hat{\bbeta}(\tau) - \bbeta_0(\tau) \}$.

\section{Additional Simulation Results}
\begin{figure}
\renewcommand{\thefigure}{S\arabic{figure}}
\centering
\includegraphics[width=16cm, height=16cm]{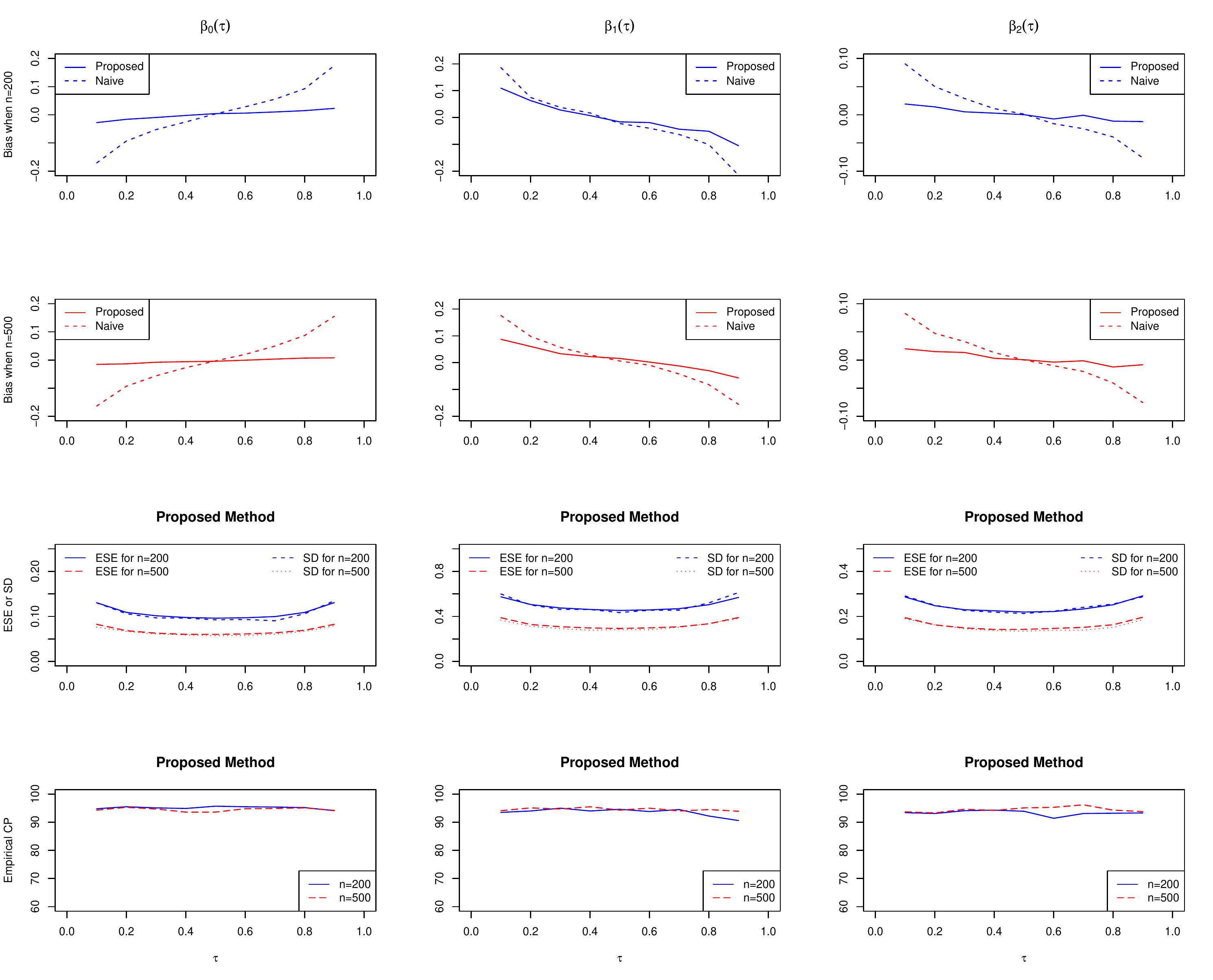}
\caption{Simulation results for Case 2. Lines in blue correspond to the results with $n=200$ and lines in red correspond to the results with $n=500$. ESE stands for the estimated standard error, SD stands for the empirical standard deviation, and CP stands for the coverage probability of a 95\% confidence interval. }  \label{NormalIndep}
\end{figure}

\begin{figure}
\renewcommand{\thefigure}{S\arabic{figure}}
\centering
\includegraphics[width=16cm, height=16cm]{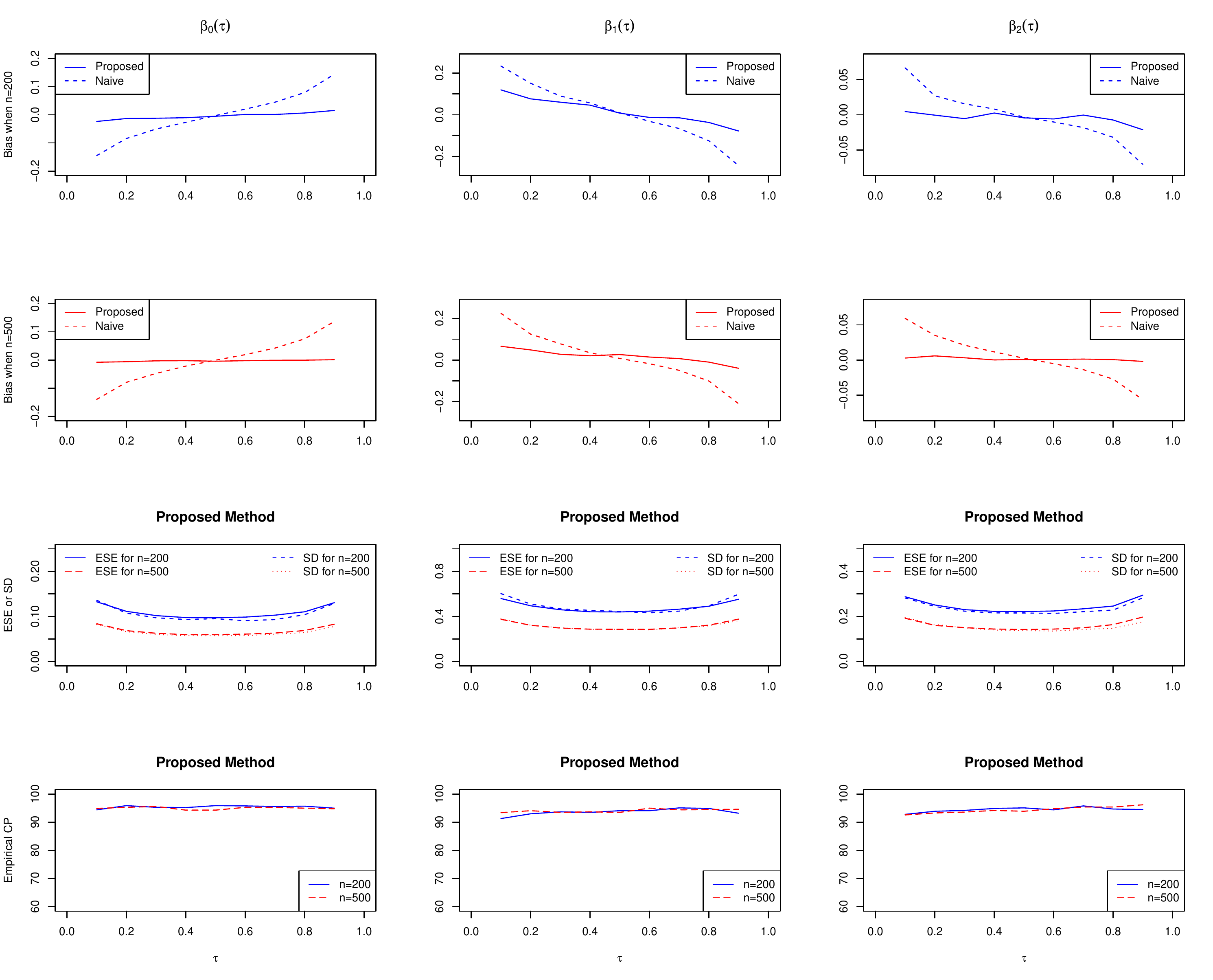}
\caption{Simulation results for Case 3. % where the errors are generated from Laplace distribution and they are dependent of the covariates.
Lines in blue correspond to the results with $n=200$ and lines in red correspond to the results with $n=500$. ESE stands for the estimated standard error, SD stands for the empirical standard deviation, and CP stands for the coverage probability of a 95\% confidence interval.}
\label{fig2}
\end{figure}

\begin{figure}
\renewcommand{\thefigure}{S\arabic{figure}}
\centering
\includegraphics[width=16cm, height=16cm]{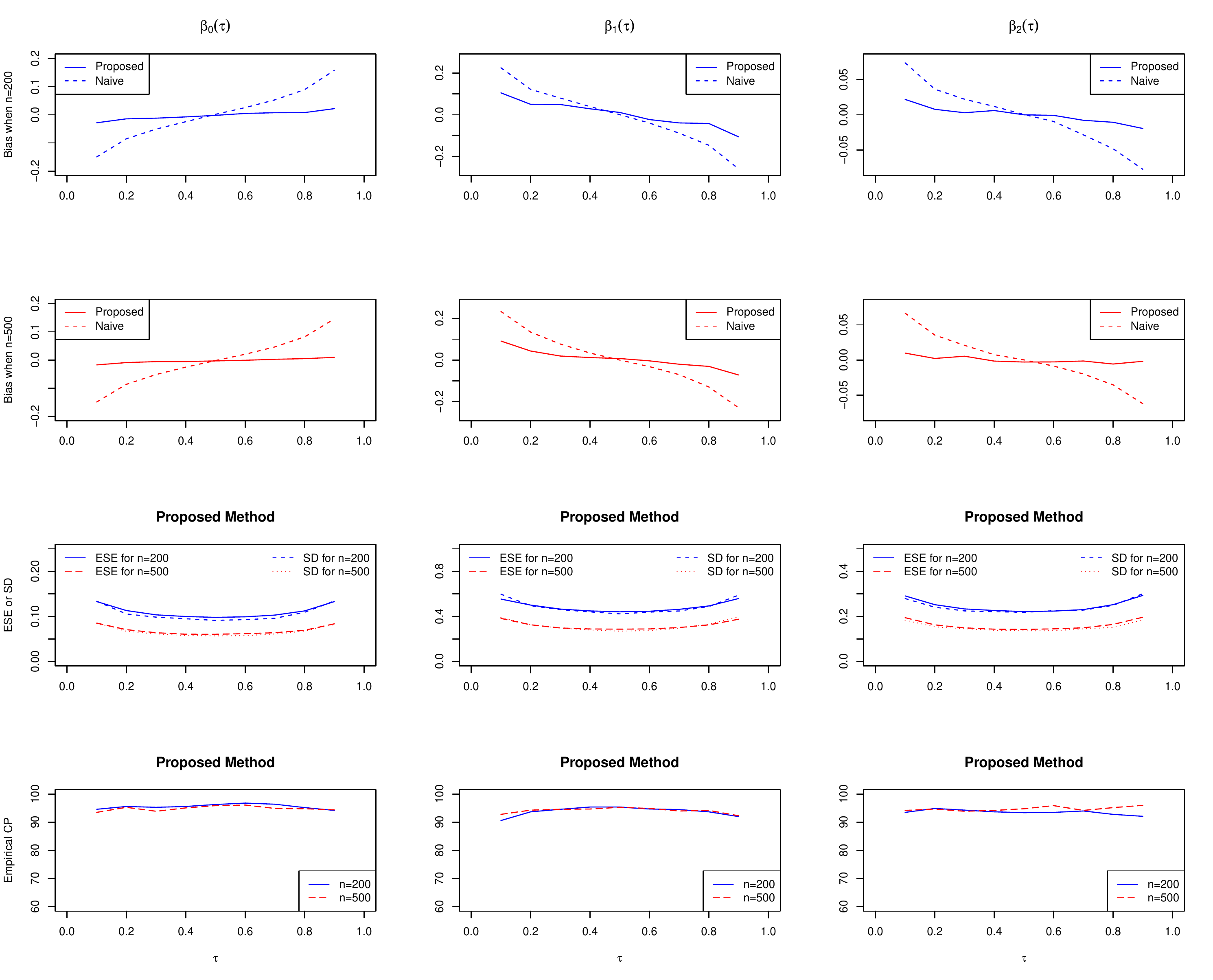}
\caption{Simulation results for Case 4. Lines in blue correspond to the results with $n=200$ and lines in red correspond to the results with $n=500$. ESE stands for the estimated standard error, SD stands for the empirical standard deviation, and CP stands for the coverage probability of a 95\% confidence interval.  }  \label{NormalDep}
\end{figure}

\begin{figure}
\renewcommand{\thefigure}{S\arabic{figure}}
\centering
\includegraphics[width=16cm, height=16cm]{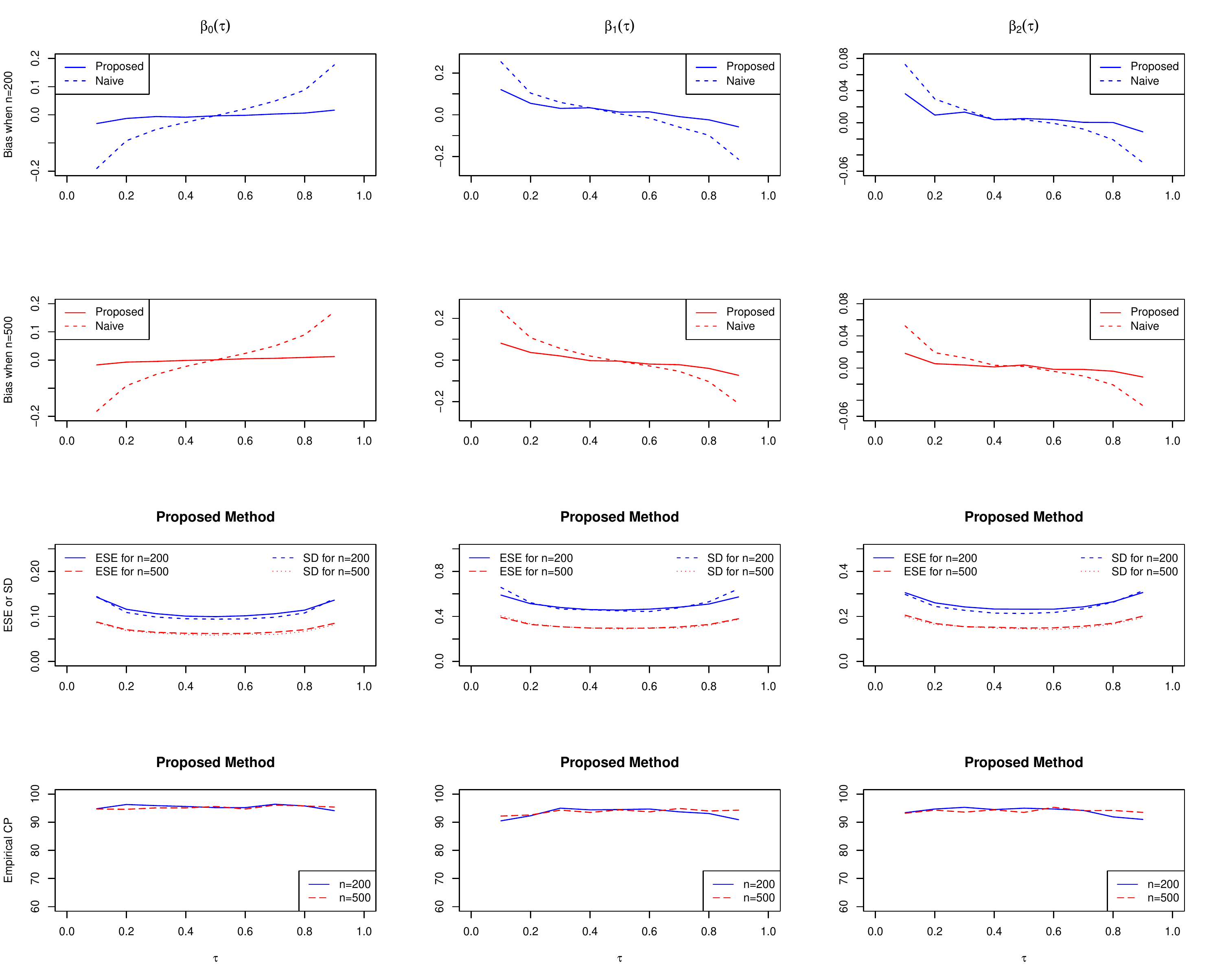}
\caption{Simulation results for quadratic trajectory model with Laplace errors. Lines in blue correspond to the results with $n=200$ and lines in red correspond to the results with $n=500$. ESE stands for the estimated standard error, SD stands for the empirical standard deviation, and CP stands for the coverage probability of a 95\% confidence interval. }  \label{NormalDep2}
\end{figure}

In this Section, we present some simulation results not included in the main manuscript. Fig. \ref{NormalIndep}, Fig. \ref{fig2} and Fig. \ref{NormalDep} present the simulation results for Cases 2, 3, and 4 respectively. The simulation results for the quadratic trajectory model with Normal errors are given in Fig. \ref{NormalDep2}.

\section{Linear Trajectory Model for DURABLE Data}

In this Section, we analyze the DURABLE data using the linear trajectory model for longitudinal measurements of HbA1c and quantile regression model for the corresponding decreasing rate.

Let $Y_{ij}$ represent the $j$-th HbA1c measurements of the $i$th individual recorded in the $t_{ij}$ week since the study enrollment ($j=1,\ldots, m_i$). We assume a linear trajectory model for within-subject HbA1c  measurements during the 24 week follow-up period. That is,
$$
Y_{ij} = \alpha_{i0} + \alpha_{i1} t_{ij} +\epsilon_{ij}.
$$
Under this model, the random intercept $\alpha_{i0}$ denotes the subject-specific baseline HbA1c measurement, and the subject-specific decreasing rate of HbA1c at every time point is given by  $-\alpha_{i1} $. In this analysis, we set $B_i\doteq -\alpha_{i1} $, and exploit how treatments and other risk factors influence the HbA1c reduction rate after the initiation of the assigned insulin treatment via quantile regression.

\renewcommand{\thetable}{S\arabic{table}}
\begin{table}\caption{ Summary statistics for the DURABLE data }
\centering
{\footnotesize
\fbox{
\begin{tabular}{ccccc}
           &   \multicolumn{4}{c}{Category by therapy and SU use}   \\ [1mm]
    \cline{2-5}   \\ [-2mm]
    &    GL       &     LM75/25     &   GL      &  LM75/25        \\ [1mm]
    &  no SU      &    no  SU       &   SU      &    SU           \\
\hline
  $n$      &  81            &    76         &   880              &      853              \\ [1mm]
 (\%)      & (4.29 \%)      & (4.02 \%)     &  (46.56 \%)        &    (45.13 \%)          \\
\end{tabular}  }   }\label{tableS1}
\end{table}

The covariates considered in this analysis is the same as in the quadratic trajectory model, including {\it therapy}, {\it sulfouse}, {\it basfglu},  {\it basfins}, and {\it therapy}$*${\it sulfouse}.  We exclude subjects with only one HbA1c measurement. This leads to a sample size of 1,890 subjects for this data analysis. The continuous covariate {\it basfglu} ranges from 0.23 to 26, with mean=11 and standard deviation=3.7. The {\it basfins} ranges from $-2$ to 143, with mean=10 and standard deviation=9.2. Table S1 % \ref{tableS1}
presents other summary statistics for the discrete covariates {\it therapy} and {\it sulfouse}.

\begin{figure}
\renewcommand{\thefigure}{S\arabic{figure}}
\centering
\includegraphics[width=16cm, height=16cm]{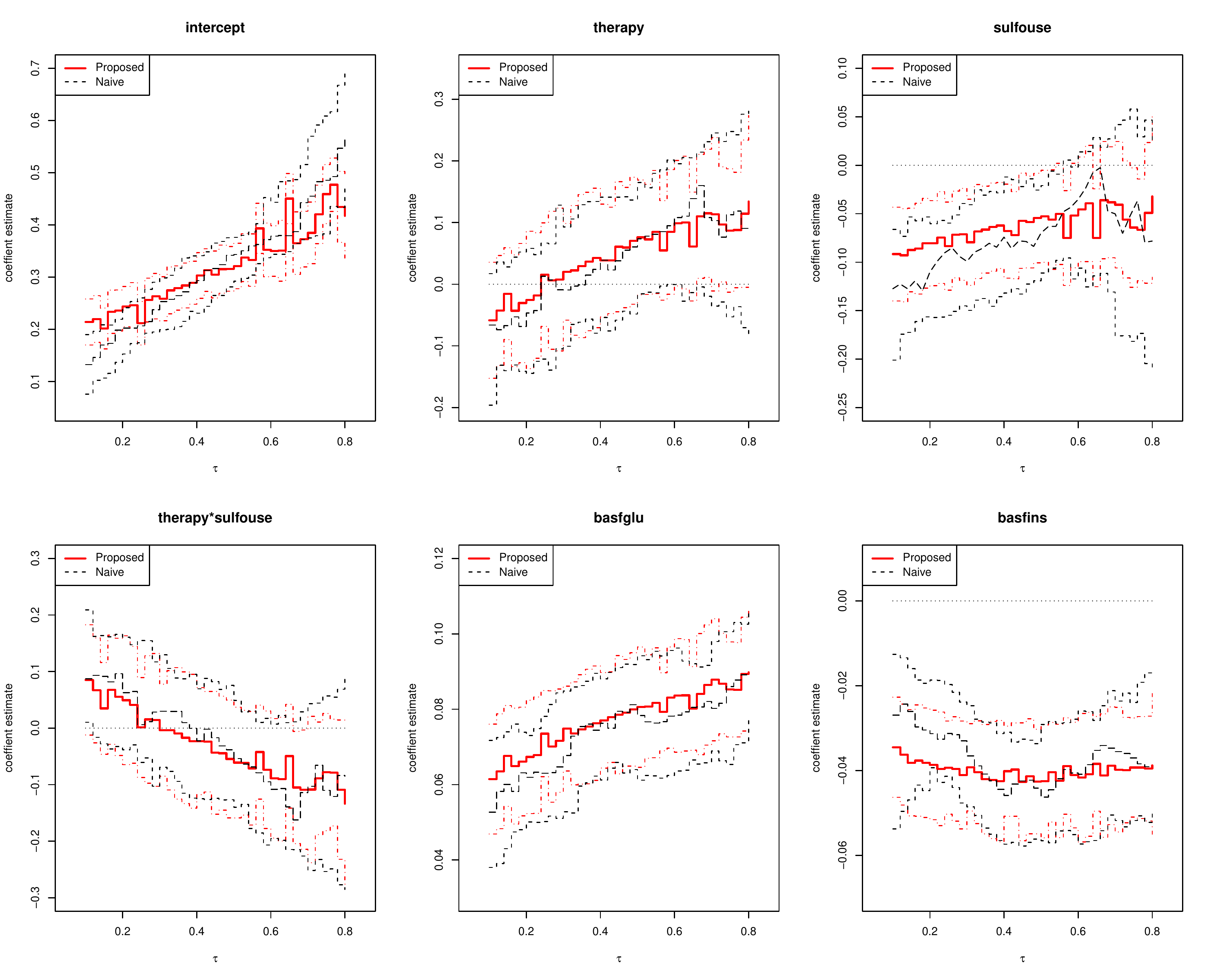}
\caption{The DURABLE data example: the proposed coefficient estimates (red solid line) and the 95\% pointwise confidence intervals (red dot-dashed line), the naive coefficient estimates (black long-dashed line) and 95\% pointwise confidence intervals (black dashed line) }
 \label{figs5}
\end{figure}

\begin{figure}
\renewcommand{\thefigure}{S\arabic{figure}}
\centering
\includegraphics[width=16cm, height=8cm]{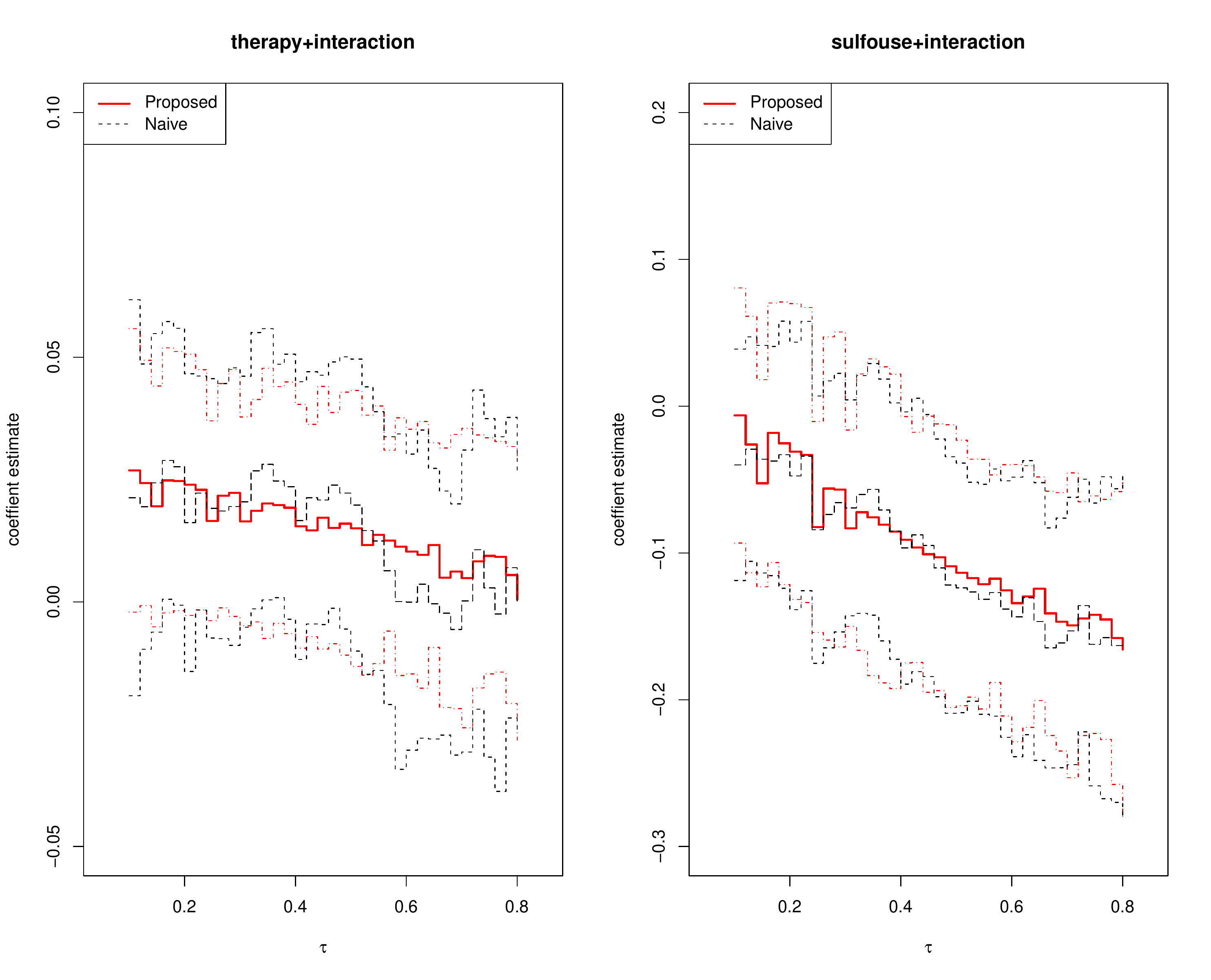}
\caption{The DURABLE data example: the estimates for {\it therapy} coefficient plus {\it therapy}*{\it sulfouse} coefficient and {\it sulfouse} coefficient plus {\it therapy}*{\it sulfouse} coefficient, and the corresponding 95\% pointwise confidence intervals. Red lines correspond to the proposed method and black lines correspond to the naive method. }
\label{figs6}
\end{figure}

We apply the proposed method to perform quantile regression for $B_i$ on the covariates specified above at quantile levels equally spaced between $0.1$ and $0.8$ with step size $0.02$. For selecting the smoothing parameter $h$, we employ the procedure introduced in Subsection 3.4 on a $h$-grid between $0.8$ and $1.5$ with step size 0.1.
In Fig. \ref{figs5}, we plot the proposed estimated coefficients (red solid line) along with the 95\% pointwise confidence intervals (red dot dashed lines) for $\tau \in [0.1, 0.8]$ based on 200 resampling samples. The naive estimators (black dashed lines) are also plotted for comparisons.
In Fig. \ref{figs6}, we  plot the estimates for {\it therapy} coefficient $+$ {\it therapy}*{\it sulfouse} coefficient and {\it sulfouse} coefficient $+$ {\it therapy}*{\it sulfouse} coefficient, which represent the treatment effect for subjects with SU use and the effect of SU use for subjects receiving LM75/25.
The interpretations of the results in Fig. \ref{figs5} and Fig. \ref{figs6} are very consistent with the results obtained based on a quadratic trajectory model.  This suggests the robustness of the proposed trajectory quantile regression method to the specification of the underlying trajectory model.

%%%%%%%%%%%%%%%%%%%%%%%%%%%%%%%%%%%%%%%%%%%%%%%%%%%%%%%%%%%%%%%%%%%%%%%%%%%%%%%%%%%%%%%%%%%%%%%%%%%%%%%%%%%%%%%%%%%%%%%%%%%%%
%\markboth{\hfill{\footnotesize\rm Huijuan Ma, Limin Peng and Haoda Fu} \hfill}
%{\hfill {\footnotesize\rm Quantile Regression of Latent Longitudinal Trajectory Features} \hfill}
%
%%\iffalse
%\bibhang=1.7pc
%\bibsep=2pt
%\fontsize{9}{14pt plus.8pt minus .6pt}\selectfont
%\renewcommand\bibname{\large \bf References}
%%\begin{thebibliography}{11}
%\expandafter\ifx\csname
%natexlab\endcsname\relax\def\natexlab#1{#1}\fi
%\expandafter\ifx\csname url\endcsname\relax
%  \def\url#1{\texttt{#1}}\fi
%\expandafter\ifx\csname urlprefix\endcsname\relax\def\urlprefix{URL}\fi
%%\fi

\bibliographystyle{biom}      % Chicago style, author-year citations
\bibliography{longitudinal}   % name your BibTeX data base